\documentclass[sigconf,screen]{acmart}

\setcopyright{none}
\renewcommand\footnotetextcopyrightpermission[1]{}
\acmConference[Preprint]{Preprint}{xxxx/xx/xx -- xxxx/xx/xx}{yyyy}
\acmYear{2027}
\acmDOI{}
\acmISBN{}

\usepackage{booktabs}
\usepackage{tikz}
\usepackage{pgfplots}
\pgfplotsset{compat=1.18}
\usetikzlibrary{positioning,calc,shapes.geometric,decorations.pathreplacing}
\usepackage[capitalise,nameinlink]{cleveref}
\makeatletter
\let\mrg@orig@mkteasers\@mkteasers
\def\@mkteasers{\mrg@orig@mkteasers
\ifx\@abstract\@lempty\else
\global\setbox\mktitle@bx=\vbox{\noindent\unvbox\mktitle@bx\par
	\vskip 2ex
	{\centering\small\bfseries\abstractname\par}%
	\vskip 1ex
	{\small\leftskip=3em\rightskip=3em\parindent=1em\noindent\ignorespaces\@abstract\par}%
	\vskip 2ex}%
\fi}
\def\@mkabstract{}
\makeatother

\AtEndPreamble{%
\theoremstyle{acmdefinition}
\newtheorem{remark}[theorem]{Remark}

}

\newcommand{\MRG}{\textsf{MRG}}
\newcommand{\RDO}{\textsf{RDO}}
\newcommand{\E}{\mathbb{E}}
\newcommand{\Prob}{\Pr}
\newcommand{\eps}{\varepsilon}

\newcommand{\cJ}{\mathcal{J}}

\newcommand{\glanceref}{\cref{sec:glance}}

\begin{document}
\setlength{\emergencystretch}{2.5em}
\settopmatter{printacmref=false, printccs=false, printfolios=true}
\title{Fluid Limits and Upper Bounds for Randomized Greedy Matching}

\author{Sayan Mukherjee}
\affiliation{%
	\institution{Department of Physics, Graduate School of Sciences, The University of Tokyo}
	\city{Hongo 7-3-1, Bunkyo-ku, Tokyo}
	\country{Japan 113-0033}}
\email{sayan@phys.s.u-tokyo.ac.jp}

\begin{abstract}
	The Modified Randomized Greedy (\MRG) algorithm~\cite{Tin84, ADFS95} is a classical greedy algorithm for finding approximate maximum matchings.
	\MRG{} repeatedly picks a uniformly random unmatched vertex and matches it to a uniformly random unmatched neighbor, if one exists.
	Since 1991, the best upper bound on its approximation ratio has been $2/3$, given by the Bomb graphs of Dyer and Frieze~\cite{DF91}.
	We improve this bound to $0.6569$ on general graphs and $0.6607$ on bipartite graphs.
	
	Our constructions consist of copies of small gadgets, with specified vertex classes joined completely.
	We track each algorithm by counting gadget copies with each possible configuration of remaining vertices.
	Wormald's differential equation method~\cite{Wor95, Wor99} shows that these counts, scaled by the size parameter of the construction, converge to a deterministic trajectory known as the \emph{fluid limit}, which is described by ordinary differential equations.
	This fluid limit approximation holds while the relevant vertex classes have positive density, and so we devise a separate counting argument determines the contribution from the end of the run.
	
	We then apply our framework to Random Decision Order (\RDO), in which vertices are processed in uniformly random order and choose unmatched neighbors according to fixed adversarial preferences.
	On bipartite graphs, we determine the exact limiting ratio on the Double-Bomb family of Tang, Wu and Zhang~\cite{TWZ23}.
	Its minimum is $31/48$, confirming their numerical estimate of about $0.646$.
	On general graphs, we improve their upper bound of $5/8$, given by the paw graph, to $0.5946$ on a new family.
	Finally, we give one possible explanation for why $2/3$ seemed to be optimal for \MRG{} in the concluding remarks.
\end{abstract}

\ccsdesc[500]{Theory of computation~Randomness, geometry and discrete structures}
\ccsdesc[500]{Mathematics of computing~Approximation algorithms}
\ccsdesc[500]{Mathematics of computing~Matchings and factors}
\ccsdesc[500]{Theory of computation~Approximation algorithms analysis}

\keywords{randomized greedy matching, approximation ratio, differential equation method, fluid limits}

\maketitle

\section{Introduction}
\label{sec:intro}

A matching in a graph $G$ is a set of vertex-disjoint edges, and $\nu(G)$ denotes the size of a \emph{maximum} matching.
Any \emph{maximal} matching (matching that cannot be extended by adding new edges) has at least $\nu(G)/2$ edges, and so the simplest greedy algorithm, which adds edges one at a time as long as possible, is a $1/2$-approximation.
A long line of work asks how much \emph{randomness} helps such greedy algorithms~\cite{Tin84,DF91,ADFS95,PS12,CCWZ18,GT12,BP17,TWZ23,DY26}.
Besides being natural, these algorithms are oblivious and local, i.e. they never look at the graph beyond the neighbors of the current vertex (without full knowledge of the graph).
This matters when matches must be made as requests stream in, long before a maximum matching could be computed.
In search advertising, an ad must be chosen for each query as it arrives, as in the AdWords problem of search engines such as Google~\cite{MSVV07,Meh13}, and in real-time bidding each display impression is auctioned within $10$ to $100$ milliseconds~\cite{YWZ13,YWLQ14}.
Ride-sharing platforms assign riders to vehicles in batches collected over windows of about $30$ seconds~\cite{AMSWFR17}, and randomized online assignment is a standard way to balance load in distributed systems~\cite{ABKU99}.
A greedy rule that processes vertices in random order uses only local information at each step, and its approximation ratio measures what is lost by not computing the optimum.

The most natural randomized greedy algorithm processes the \emph{edges} in a uniformly random order.
Dyer and Frieze~\cite{DF91} showed that this does not beat $1/2$ in the worst case.
Randomizing over \emph{vertices} instead goes back to Tinhofer~\cite{Tin84} (see~\cite{BP17}), and Aronson, Dyer, Frieze, and Suen~\cite{ADFS95} analyzed the following algorithm under the name Modified Randomized Greedy.

\begin{definition}[Modified Randomized Greedy, \MRG]
	Start with $M=\varnothing$ and $U=V(G)$.
	While $U\ne\varnothing$, pick $u\in U$ uniformly at random.
	If $u$ has a neighbor in $U$, pick one, $v$, uniformly at random, add $uv$ to $M$, and remove $u,v$ from $U$.
	Otherwise remove $u$ from $U$.
\end{definition}

\MRG{} is an algorithm for the \emph{randomized greedy matching} problem, also called the oblivious matching problem~\cite{GT12,TWZ23}.
The algorithm chooses the order in which it tries vertex pairs without looking at the graph, adds every tried pair that is an edge between two unmatched vertices, and its quality is defined by the worst-case ratio $\E|M|/\nu(G)$.
The point of \MRG{} is that a vertex with few available neighbors is matched along one of them with correspondingly large probability.
\subsection{Background and Related Work}
\label{sec:related}

Aronson, Dyer, Frieze, and Suen~\cite{ADFS95} proved that \MRG{} is $(1/2+1/400{,}000)$-approximate, the first randomized greedy algorithm to provably beat $1/2$ on general graphs.
Poloczek and Szegedy~\cite{PS12} claimed an improvement to $1/2+1/256$, and later work pointed out gaps in their proof~\cite{CCWZ18,TWZ23}.
Tang, Wu, and Zhang~\cite{TWZ23} proved the current best lower bound of $0.531$, by analyzing the weaker algorithm \RDO{} defined below, and Derakhshan and Yu~\cite{DY26} recently gave a unified framework for these analyses.

\begin{figure*}[htbp]
	\centering
	\begin{tikzpicture}[
		cls/.style={circle, draw=black!70, fill=gray!8, thick, minimum size=0.8cm, inner sep=0pt},
		ant/.style={circle, draw=black!70, fill=white, thick, minimum size=0.8cm, inner sep=0pt},
		sat/.style={circle, draw=black!70, fill=white, thick, minimum size=0.6cm, inner sep=0pt},
		blk/.style={line width=3.2pt, draw=gray!55},
		opt/.style={thick, dashed, draw=orange!80!black},
		sz/.style={font=\normalsize},   
		lab/.style={font=\normalsize, align=center}]
		\begin{scope}
			\node[cls] (X) at (0,0) {$X$};
			\node[cls] (Y) at (1.5,0) {$Y$};
			\node[ant] (Xp) at (0,1.5) {$X'$};
			\node[ant] (Yp) at (1.5,1.5) {$Y'$};
			\draw[blk] (X) -- (Y);
			\draw[opt] (X) -- (Xp); \draw[opt] (Y) -- (Yp);
			\node[lab] at (0.75,-2.75) {(a) the bomb graph\\ $B_n$};
		\end{scope}
		\begin{scope}[xshift=3.5cm]
			\node[cls] (X) at (0,0) {$X$};
			\node[cls] (Y) at (1.5,0) {$Y$};
			\node[ant] (Xp) at (0,1.5) {$X'$};
			\node[ant] (Yp) at (1.5,1.5) {$Y'$};
			\node[sat] (P) at (0,-1.5) {$P$};
			\node[sat] (Q) at (1.5,-1.5) {$Q$};
			\node[sz] at (0,-2.05) {$|P|=\sigma n$};
			\node[sz] at (1.5,-2.05) {$|Q|=\sigma n$};
			\draw[blk] (X) -- (Y); \draw[blk] (X) -- (P); \draw[blk] (Y) -- (Q);
			\draw[opt] (X) -- (Xp); \draw[opt] (Y) -- (Yp); \draw[opt] (P) -- (Q);
			\node[lab] at (0.75,-2.75) {(b) the Double-Bomb\\ $S_n(\sigma)$};
		\end{scope}
		\begin{scope}[xshift=7.7cm]
			\node[cls] (X) at (0,0) {$X$};
			\node[cls] (Y) at (1.5,0) {$Y$};
			\node[ant] (Xp) at (0,1.5) {$X'$};
			\node[ant] (Yp) at (1.5,1.5) {$Y'$};
			\node[sat] (P) at (0,-1.5) {$P$};
			\node[sat] (Q) at (1.5,-1.5) {$Q$};
			\node[sz] at (0,-2.05) {$|P|=\sigma n$};
			\node[sz] at (1.5,-2.05) {$|Q|=\sigma n$};
			\draw[blk] (X) -- (Y); \draw[blk] (X) -- (P); \draw[blk] (Y) -- (Q);
			\draw[blk] (X) to[out=200,in=160,looseness=7] (X);
			\draw[blk] (Y) to[out=20,in=-20,looseness=7] (Y);
			\draw[opt] (X) -- (Xp); \draw[opt] (Y) -- (Yp); \draw[opt] (P) -- (Q);
			\node[lab] at (0.75,-2.75) {(c) the Double-Bomb with cliques\\ $S^K_n(\sigma)$};
		\end{scope}
		\begin{scope}[xshift=12.0cm]
			\node[ant] (A) at (0,1.5) {$A$};
			\node[ant] (B) at (1.5,1.5) {$B$};
			\node[ant] (C) at (0.75,0.2) {$C$};
			\node[ant] (D) at (0.75,-1.3) {$D$};
			\node[sat, fill=gray!8] (E) at (2.25,-1.3) {$E$};
			\node[sat] (F) at (3.6,-1.3) {$F$};
			\node[sz] at (2.92,-1.85) {$|E|=|F|=\beta n$};
			\draw[thick] (A) -- (C); \draw[thick] (B) -- (C);
			\draw[opt] (A) -- (B); \draw[opt] (C) -- (D); \draw[opt] (E) -- (F);
			\draw[blk] (D) -- (E);
			\draw[blk] (E) to[out=60,in=120,looseness=7] (E);
			\node[lab] at (1.8,-2.75) {(d) the paw with a pool\\ $PP_n(\beta)$};
		\end{scope}
	\end{tikzpicture}
	\Description{Four class-level diagrams: the bomb graph, the Double-Bomb, the Double-Bomb with cliques on both hub sides, and copies of the paw attached to a pool of edges whose one side is a clique.}
	\caption{The four constructions, drawn at the level of classes. Each circle is a class of $n$ vertices unless its size is shown. Thick gray bands are complete joins between two classes, and a gray loop is a clique on a class. Dashed lines are perfect matchings between two classes; together they form the optimal perfect matching $M^*$. (a) The bomb graph. (b) The Double-Bomb~\cite[Figure~8]{TWZ23}: the satellite $p_j\in P$ sees all of $X$, and its partner $q_j\in Q$ sees all of $Y$. (c) The Double-Bomb with cliques. (d) Copies of the paw $abc$ plus $cd$ (thin lines are edges inside each copy), and a pool of edges $ef$ whose $E$-side is a clique joined to every $d$.}
	\label{fig:constructions}
\end{figure*}
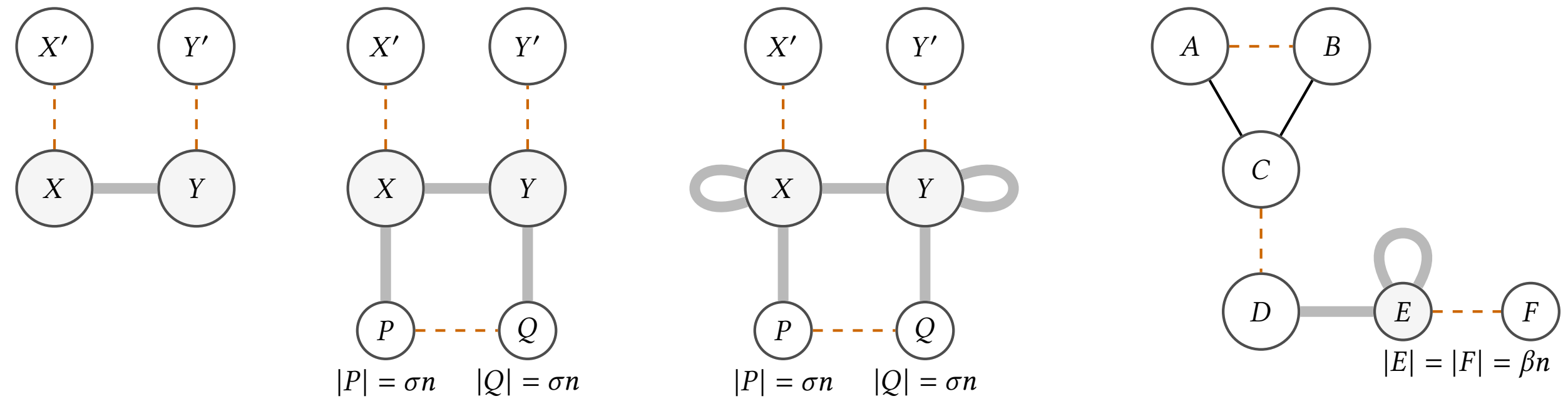

Despite substantial interest and progress on the lower bound, the upper bound of $2/3$ has not moved since 1991.
The \emph{bomb graph} $B_n$ of Dyer and Frieze~\cite{DF91} consists of a complete bipartite graph $K_{n,n}$ between parts $X$ and $Y$, with a pendant vertex (an \emph{antenna}) attached to each of its $2n$ vertices; see \cref{fig:constructions}(a).
We call $X$ and $Y$, the ``hubs" of $B_n$.
The antenna edges form a perfect matching, but the expected size of the matching found by \MRG{} is only a $2/3+o(1)$ fraction of it.
It has been widely believed that bomb graphs are the worst case, that is, that \MRG{} is exactly $2/3$-approximate (see, e.g.,~\cite{PS12}).
Deligkas, Mertzios, and Spirakis~\cite{DMS17} even reduced the question to a weighted problem: they showed that if their randomized approximation algorithm (RGMA) for greedy matching on weighted ``bush graphs'' is $\rho$-approximate, then \MRG{} is $(\rho-\eps)$-approximate for every $\eps>0$, and they conjectured that the tight value of $\rho$ is $2/3$.
The remaining gap $[0.531,\,2/3]$ is perhaps one of the basic open problems about randomized greedy matching.

Another related algorithm in this area is the Random Decision Order (\RDO{}) algorithm of Tang, Wu, and Zhang~\cite{TWZ23}.
The idea is to isolate the role of the two sources of randomness (picking an unmatched vertex, and that vertex picking an unmatched neighbor) in \MRG.
In \RDO, the next vertex is chosen uniformly at random, as in \MRG, but each vertex matches its \emph{favorite} available neighbor according to a fixed, adversarially chosen preference order.
They proved that \RDO{} is at least $0.639$-approximate on bipartite graphs and $0.531$-approximate on general graphs.
Since every preference order is allowed, both lower bounds transfer to \MRG.
For upper bounds, they showed that \RDO{} is at most $5/8$-approximate on general graphs, using a four-vertex graph, and they reported an \emph{experimental} upper bound of about $0.646$ on bipartite graphs, using a graph they call the Double-Bomb.
They left the analysis of this graph as future work.

The Double-Bomb $S_n(\sigma)$ of Tang, Wu, and Zhang~\cite[Appendix~B.1]{TWZ23} is drawn in \cref{fig:constructions}(b).
It is constructed from the bomb graph with ``hubs'' $X$ and $Y$ and ``antennas'' $X'$ and $Y'$, by adding sets $P$ and $Q$ such that $|P|=|Q|=\sigma n$.
We call $P$ and $Q$ ``satellites''; every satellite of $P$ sees all of $X$, and every satellite of $Q$ sees all of $Y$.
Finally, we add a matching of size $\sigma n$ between $P$ and $Q$.

\paragraph{Vertex-iterative greedy algorithms:}
Following~\cite{DY26}, these algorithms are described by a decision order on the vertices and a preference order for each vertex; \cref{tab:models} lists the ones we mention.
For Ranking on general graphs, the lower bound improved from $0.523$~\cite{CCWZ18} to $0.526823$~\cite{CCW18}, then to $0.5469$~\cite{DRSY26}, and recently to $0.560$~\cite{DY26}.
For its upper bound, Karande, Mehta, and Tripathi~\cite{KMT11} constructed a hard instance on which Ranking achieves at most $0.727$; Chan, Chen, Wu, and Zhao~\cite{CCWZ18} modified it and found about $0.724$ by experiment, and the Double-Bomb of Tang, Wu, and Zhang~\cite{TWZ23} is modeled on their instance.
On bipartite graphs with random arrivals, Mahdian and Yan~\cite{MY11} analyzed Ranking~\cite{KVV90} with strongly factor-revealing LPs.
For the oblivious matching problem itself, Goel and Tripathi~\cite{GT12} showed that no algorithm is better than about $0.792$-approximate, and Huang, Sun, Wu, and Zhao~\cite{HSWZ25} recently improved this bound for general graphs to about $0.7583$, and also proved a bound of $0.7961$ for bipartite graphs; these bounds hold for every oblivious algorithm and are much weaker than our bounds for the specific algorithms \MRG{} and \RDO.
In the fully online model, where all vertices arrive and depart online, Huang, Tang, Wu, and Zhang~\cite{HTWZ26} recently showed that a fractional Water-Filling algorithm is exactly $(2-\sqrt2)$-competitive, improved this to $0.599$, and lowered the hardness of fractional fully online matching to $0.6132$; there the adversary also controls when vertices arrive and depart, so their bounds and ours do not imply each other.

\begin{table}[t]
	\caption{Vertex-iterative greedy algorithms, following~\cite{DY26}. In UUR the common preference order is independent of the decision order; in Ranking the two coincide.}
	\label{tab:models}
	\small
	\begin{tabular}{lll}
		\toprule
		Algorithm & Decision order & Preference orders \\
		\midrule
		\MRG & uniform random & independent uniform, per vertex \\
		\RDO & uniform random & fixed, adversarial \\
		Ranking & uniform random & the decision order itself \\
		UUR & uniform random & one common uniform order \\
		FRanking & adversarial & one common uniform order \\
		\bottomrule
	\end{tabular}
\end{table}

\paragraph{Differential equation methods:}

Differential equation approximations of Markov chains go back to Kurtz~\cite{Kur70}; see Darling and Norris~\cite{DN08} for a survey.
In combinatorics, Wormald's method~\cite{Wor95,Wor99,War19} is the standard tool for analyzing random greedy processes, and differential equations have been used for greedy matching on random graphs since Karp and Sipser~\cite{KS81}.
Closest to our setting, Mastin and Jaillet~\cite{MJ13} used Wormald's theorem to analyze greedy online matching on the random bipartite graph $G(n,n,p)$, in what they describe as one of the simplest applications of the theorem.
Computer search and computer-aided analysis also appear in the hardness results of~\cite{CCWZ18,TWZ23,HSWZ25}.
We describe the role of computation in this work in \cref{sec:toolkit-search}.

\subsection{Our Results}

In this work, we give the first improvement on the $2/3$ upper bound for \MRG, both on general and on bipartite graphs, and we improve the upper bounds for \RDO.
\Cref{tab:results} summarizes our results.

\begin{table*}[htbp]
	\caption{Approximation ratio of \MRG{} and \RDO. The upper bounds in the last column are proved in this paper.}
	\label{tab:results}
	\small
	\begin{tabular}{lccc}
		\toprule
		& Lower bound & Previous upper bound & Our upper bound \\
		\midrule
		\MRG, general graphs & $0.531$~\cite{TWZ23} & $2/3$~\cite{DF91} & $0.6569$ (\cref{thm:mrg-general}) \\
		\MRG, bipartite graphs & $0.639$~\cite{TWZ23} & $2/3$~\cite{DF91} & $0.6607$ (\cref{thm:mrg-bip}) \\
		\RDO, bipartite graphs & $0.639$~\cite{TWZ23} & $\approx 0.646$, experimental~\cite{TWZ23} & $31/48 = 0.64583\ldots$ (\cref{thm:rdo-bip}) \\
		\RDO, general graphs & $0.531$~\cite{TWZ23} & $5/8 = 0.625$~\cite{TWZ23} & $0.5946$ (\cref{thm:rdo-general}) \\
		\bottomrule
	\end{tabular}
\end{table*}

We first determine the limiting approximation ratio of \RDO{} on the Double-Bomb family $S_n(\sigma)$ recalled above.

\begin{theorem}[\RDO{} on bipartite graphs]
	\label{thm:rdo-bip}
	On the Double-Bomb $S_n(\sigma)$ with the preferences of Tang, Wu, and Zhang, the expected ratio $\E|M|/\nu$ of \RDO{} converges to
	\[
	\rho_{\RDO}(\sigma)=\frac23-\frac23\,w^3+w^4,\text{ where } w=\sqrt{\frac{\sigma}{2+\sigma}}.
	\]
	In particular, the minimum over $\sigma$ is attained at $\sigma=2/3$ and equals $31/48$, so \RDO{} is at most $31/48$-approximate on bipartite graphs.
\end{theorem}

The minimizer $\sigma=2/3$ corresponds to the ratio $3/2$ between the two part sizes in~\cite[Appendix~B.1]{TWZ23}, which is exactly the shape that Tang, Wu, and Zhang found to be optimal in their computational experiments.
The same family also beats $2/3$ for \MRG{} when $\sigma=5/4$.

\begin{theorem}[\MRG{} on bipartite graphs]
	\label{thm:mrg-bip}
	As $n\to\infty$, the expected ratio of \MRG{} on $S_n(5/4)$ converges to an explicit integral $\rho_S(5/4)=0.66068\ldots$.
	In particular, \MRG{} is at most $0.6607$-approximate on bipartite graphs.
\end{theorem}

Next, by turning the hubs $X$ and $Y$ into cliques in $S_n(\sigma)$, we obtain a non-bipartite graph family $S^K_n(\sigma)$.
We call these the \emph{Double-Bomb with cliques} (\cref{fig:constructions}(c)), and prove that they are even harder for \MRG{} on general graphs.

\begin{theorem}[\MRG{} on general graphs]
	\label{thm:mrg-general}
	As $n\to\infty$, the expected ratio of \MRG{} on $S^K_n(1)$ converges to an explicit integral $\rho_K(1)=0.65682\ldots$.
	In particular, \MRG{} is at most $0.6569$-approximate on general graphs.
\end{theorem}

The constants $\rho_S$ and $\rho_K$ are one-dimensional integrals of elementary functions (\cref{sec:mrg}), which can be evaluated numerically.
By the reduction of Deligkas, Mertzios, and Spirakis~\cite{DMS17}, \cref{thm:mrg-general} also implies that RGMA on bush graphs is at most $0.6569$-approximate, so our results imply that their conjectured tight value $2/3$ is not correct.

Finally, we improve on the upper bound of $5/8$ for \RDO{} on general graphs.
We start with the \emph{paw} construction of~\cite[Theorem~B.1]{TWZ23} that achieves $5/8$ under the common preference order $c\succ b\succ a\succ d$: vertex subsets $A, B, C, D$ of size $n$ with perfect matchings between parts $\{A,B\},\{B,C\},\{C,A\},\{C,D\}$.
Then, we add two sets $E$ and $F$ with $|E|=|F|=\beta n$ which are joined themselves by a perfect matching, $E$ is a clique, and every vertex of $E$ sees every vertex of $D$ (See \cref{fig:constructions}(d)).
Analyzing \RDO{} on this family $PP_n(\beta)$ with a suitable common preference order and selecting a suitable $\beta$ gives us the following theorem.

\begin{theorem}[\RDO{} on general graphs]
	\label{thm:rdo-general}
	There are graphs $G_n$ and preference orders for which the expected ratio of \RDO{} converges to
	\[
	\min_{0<w<1}\ \frac{2w^2+\tfrac12 w^6+\tfrac23(1-w^3)}{3w^2+1} = 0.594531\ldots,
	\]
	so \RDO{} is at most $0.5946$-approximate on general graphs.
\end{theorem}

\Cref{thm:rdo-general} improves the $5/8$ bound of~\cite{TWZ23}, and it widens their separation between \RDO{} on bipartite graphs (at least $0.639$) and on general graphs (now at most $0.5946$).

\paragraph{Organization:}
In \cref{sec:prelims}, we give the formal background for the analysis, including the continuous time formulation, templates, and Wormald's theorem.
In \cref{sec:primer}, we introduce the toolkit, illustrate it on the Double-Bomb, and then state and prove our certification lemma and apply it to the classical bomb graph.
We also briefly describe the role of the computational experiments.
\Cref{sec:rdo} proves \cref{thm:rdo-bip,thm:rdo-general} for \RDO{}, and \cref{sec:mrg} proves \cref{thm:mrg-bip,thm:mrg-general} for \MRG{}.
We end in \cref{sec:conclusion} with the behavior of the algorithms for graphs near the bomb graph (giving one possible explanation for why improving examples were difficult to notice), numerical evidence for smaller constants, and further questions.

\section{Preliminaries and Notation}
\label{sec:prelims}

All graphs in this work are finite and simple.
For a graph $G$, $\nu(G)$ is the size of a maximum matching, and for an algorithm $\mathcal{A}$, $M=M_{\mathcal{A}}(G)$ is its (random) output.
The \emph{approximation ratio} of $\mathcal{A}$ on a class of graphs is $\inf_G \E|M_{\mathcal{A}}(G)|/\nu(G)$ over the graphs $G$ in the class with $\nu(G)>0$, and an upper bound on it is shown by exhibiting a sequence of graphs $G_n$ with $\limsup_n \E|M(G_n)|/\nu(G_n)$ small.
A vertex $v$ is \emph{covered} by a matching $M$ if some edge of $M$ contains $v$.

\subsection{The Algorithms in Step Form and Rank Form}

Both \MRG{} and \RDO{} are classically written in \emph{step form}.
Keep a set $U$ of \emph{active} vertices, initially $V(G)$.
At each step, a uniformly random active vertex $u$ \emph{acts}.
If $u$ has an active neighbor, it chooses one, say $v$, according to its rule (uniformly random or preference), $uv$ is added to $M$ and both are removed from $U$.
If $u$ has no active neighbors it is removed from $U$.
In \MRG{} the rule is a uniformly random active neighbor, and in \RDO{} it is the first active neighbor in the fixed preference order $\succ_u$ of $u$.
Each vertex acts at most once, and in \MRG{} its neighbor is chosen uniformly conditional on the current active set.
The run on $G$ from the moment the active set is $U$ onwards is the same algorithm run on the induced subgraph $G[U]$.
Note that both algorithms output a \emph{maximal} matching, since a vertex leaves $U$ unmatched only when it has no active neighbor.

The step form has an equivalent \emph{rank form}~\cite{TWZ23}.
Every vertex independently draws a uniform \emph{rank} $y_u\in[0,1]$, the vertices are processed in increasing order of rank, and a processed vertex that is still unmatched acts as above.
To see the equivalence, let us reveal the random order only as vertices are processed.
Then, conditional on the revealed prefix and the neighbor choices made so far, the relative order of the remaining vertices is still uniform.
Removing vertices that were matched as neighbors does not reveal their unused ranks, so the first remaining active vertex is uniform in the active set.
The neighbor-selection rule is the same in both forms (uniformly for \MRG, and preference-wise for \RDO).

\subsection{The Continuous Time Formulation}
\label{sec:activation}

The rank form specifies an order, but we are free to choose a time scale for that order.
For every vertex $v$, let us set $T_v=-\log(1-y_v)$.
Since this transformation is increasing, processing vertices in increasing order of $T_v$ gives exactly the same run as processing them in increasing order of $y_v$.
Moreover, the variables $T_v$ are independent exponentials of mean $1$, since $\Prob(T_v>s)=e^{-s}$ for $s\ge 0$.
We call $T_v$ the \emph{activation time} of $v$, and ignore it if $v$ has already been matched.

The useful property in this reformulation is memorylessness.
For $\delta\ge 0$, $\Prob(T_v>s+\delta\mid T_v>s)=e^{-\delta}$.
Conditional on the run up to time $s$, the unused activation times of active vertices, measured from $s$, remain independent exponentials of mean $1$, i.e., the run has only tested whether they exceed $s$, and neighbor choices do not inspect them.
Thus each active vertex acts in the next interval of length $\delta$ with probability $\delta+O(\delta^2)$.
Equivalently, it acts at rate $1$.
Among $|U|$ active vertices, each is equally likely to act next, and the waiting time to the next action has mean $1/|U|$.
This is another direct proof that the continuous time process has the same sequence of matching decisions as the step form.

This reformulation is standard in random greedy processes, and the properties of exponentials used above are textbook facts~\cite[Section~2.3]{Nor97}.
For example, for random sequential adsorption, the random greedy independent set process studied since Flory~\cite{Flo39}, Penrose and Sudbury~\cite[Section~2.2]{PS05} explain the equivalent uniform and exponential arrival-time descriptions on graphs.
In our setting, if $\theta\in[0,1)$ denotes the rank threshold, then $s=-\log(1-\theta)$ is the corresponding exponential time.
We also refer to $s$ as \emph{log-time}.

The reason to use this time scale is that rates from different classes simply add.
Let $A(s)$ be the set of active vertices with at least one active neighbor at time $s$.
Each of these vertices creates one matching edge when it acts, at rate $1$, whereas an isolated active vertex creates none.
And therefore,
\begin{equation}
	\E|M|=\E\int_0^\infty |A(s)|\, d s.
	\label{eq:activation-integral}
\end{equation}

If normalized counts were to follow a deterministic trajectory, they determine the limiting edge-creation rate in this integral.
Wormald's theorem justifies this approximation on intervals where the drift is regular.
To tackle the final matching, we need an additional end-phase argument.






\subsection{Wormald's Differential Equation Method}

We use the following special case of Theorem~5.1 of~\cite{Wor99}, with deterministically bounded jumps; see also~\cite{Wor95}, \cite{War19} for a modern treatment, and~\cite{MJ13} for a short worked application to greedy matching.
Here $t$ counts steps of a process, and $\tau=t/n$ is the rescaled time.
Both time and counts are rescaled by $n$, because a step changes the counts by $O(1)$ and the run has $\Theta(n)$ steps, so $\tau$ and $z=Z/n$ are of order $1$, and the process has a chance to follow an ODE in $\tau$.
The filtration $\mathcal{F}_t$ is the history of the first $t$ steps, and $Z_t$ is adapted to it if $Z_t$ can be read off this history.

\begin{theorem}[Wormald~\cite{Wor99}]
\label{thm:wormald}
Let $(Z_t)_{t\ge 0}$ be a process in $\mathbb{R}^a$ adapted to a filtration $(\mathcal{F}_t)$, with $\|Z_t\|_\infty\le C_0n$ and $Z_0=nz_0$.
Let $D\subset\mathbb{R}^{a+1}$ be a bounded, connected, open set containing $(0,z_0)$, called the \emph{domain}.
Let $T_D$ be the first $t$ with $(t/n,Z_t/n)\notin D$, i.e. the \emph{exit time}.
Suppose that, for all $t<T_D$:
\begin{enumerate}
	\item[(W1)] (Bounded jumps) $\|Z_{t+1}-Z_t\|_\infty\le\beta$ for a constant $\beta$.
	\item[(W2)] (Drift and regularity) There is a map $f$ that is Lipschitz continuous on $D\cap\{\tau\ge 0\}$, and
	\[
	\big\|\E[Z_{t+1}-Z_t\mid\mathcal{F}_t]-f(t/n,Z_t/n)\big\|_\infty\le\lambda^\ast,
	\]
	where $\lambda^\ast=\lambda^\ast(n)\to 0$ as $n\to\infty$.
\end{enumerate}
Then $dz/d\tau=f(\tau,z)$, $z(0)=z_0$, has a unique solution in $D$, which extends to points arbitrarily close to the boundary $\partial D$.
Moreover, for every $\lambda=\lambda(n)\to 0$ with $\lambda>\lambda^\ast$ and every sufficiently large constant $C$, with probability $1-O\big((\beta/\lambda)e^{-n\lambda^3/\beta^3}\big)$,
\[
Z_t=n\,z(t/n)+O(\lambda n) \text{ uniformly for }0\le t\le\tau_\lambda n,
\]
where $\tau_\lambda$ is the supremum of the $\tau$ such that $(\tau',z(\tau'))$ stays at $\ell_\infty$-distance at least $C\lambda$ from the boundary $\partial D$ for all $\tau'\in[0,\tau]$.
\end{theorem}

Conceptually, if the process moves in bounded steps, and its average step is a nice function of where it currently is, then it follows the ODE ``average step $=$ velocity'' up to an $o(n)$ error, until it gets close to the boundary of the domain (the bounded, connected open set where the hypotheses hold).

The theorem only asks that the conditional drift be uniformly approximated by a function of time and the tracked variables.
We apply it in a stronger setting, to a bounded number of counts whose conditional expected changes depend only on these counts.
This holds for graphs built from many disjoint copies of a few small gadgets, with some classes joined completely, which we define next.


\subsection{Templates and Instances}

\begin{definition}[Template]
\label{def:template}
A \emph{template} consists of small graphs $H_1,\dots,H_k$ (the \emph{gadgets}), positive weights $w_1,\dots,w_k$, and a set $\cJ$ of \emph{joins}, each join being an unordered pair of classes defined below.
Its \emph{instance} $G_n$ consists of $\lfloor w_g n\rfloor$ disjoint copies of each $H_g$.
\footnote{Rounding due to fractional parts changes every count by $O(1)$, which \cref{lem:template} allows.}
The \emph{class} $(g,p)$ is the set of vertices at position $p$ of the copies of $H_g$.
For every join $\{c,c'\}\in\cJ$, every vertex of $c$ is joined to every vertex of $c'$; a join $\{c,c\}$ is a clique on $c$.
\end{definition}

In a blow-up in the combinatorial sense, every edge of a small graph becomes a complete bipartite graph.
In an instance, only the joins are complete.
The edges of a gadget join vertices of the same copy, so two classes that are adjacent in a gadget span a perfect matching.
The bomb graph $B_n$ is the instance of the template with the gadget $P_4=x'xyy'$ and the single join $\{x,y\}$.
The Double-Bomb $S_n(\sigma)$ is the instance of the template with three gadgets $K_2$, namely $xx'$, $yy'$ (weight $1$) and $pq$ (weight $\sigma$), and the joins $\{x,y\}$, $\{x,p\}$ and $\{y,q\}$.
The Double-Bomb with cliques $S^K_n(\sigma)$ has also the joins $\{x,x\}$ and $\{y,y\}$ in addition to $S_n(\sigma)$.
On an instance, unmatched vertices of the same class in the same \emph{local state} (that is, with the same set of unmatched vertices in their own copy) are interchangeable for \MRG.
For \RDO{} this holds only for preferences that respect the symmetry (see \cref{rem:tiers}).
This is why an instance is described by a few counts.

\section{The General Toolkit and Its Application to the Bomb Graph}
\label{sec:primer}

We now introduce the toolkit and illustrate it on the Double-Bomb.
We then state and prove the certification lemma, which combines Wormald's theorem with a separate bound for the remaining matching edges, and apply it to the bomb graph; \cref{sec:rdo,sec:mrg} use the same conditions.

\subsection{Our Toolkit}
\label{sec:toolkit-overview}
Let $M_{\mathcal A}(G)$ denote the matching obtained by a vertex-iterative greedy algorithm $\mathcal A$ on input graph $G$.
Observe that proving an upper bound on $\inf_G (\E|M_\mathcal A(G)|/\nu)$ requires proving the existence of a graph $G$ on which $\mathcal A$ is not optimal for an average over all stochastic runs of the algorithm.
Therefore, we need to understand how $\mathcal A$ loses edges from an optimal matching over the whole run.
On the graphs considered here, this can be done by following a fixed number of counts, even though the number of vertices tends to infinity.
Our toolkit turns these counts into a limiting approximation ratio in the following three steps.

\paragraph{Step 1: Describe the run by counts:}
In extremal combinatorics, extremal constructions are typically blow-ups of small patterns, possibly iterated~\cite{Pik14,BHLP16}, and hard instances for greedy matching have a similar shape.
In this work, we say that a \emph{template} consists of a few small \emph{gadgets} and a set of \emph{joins}, and its \emph{instance} takes many disjoint copies of each gadget and joins some \emph{classes} of vertices completely (see \cref{def:template}).
The bomb graph $B_n$ is an instance of the template $P_4=x'xyy'$ (or two disjoint edges $(x'x, y'y)$) with one join $xy$, and the Double-Bomb $S_n(\sigma)$ is an instance of a template with three disjoint edges $(x'x, y'y, pq)$ and three joins $px, xy, yq$.
Unlike in a blow-up, the edges of a gadget stay inside their copy, so they form perfect matchings between classes.
These matchings have edge density zero and are invisible to dense graph limits, but they carry the optimal matching.
Therefore the right limit object here is not a graphon but a \emph{fluid limit} of class-level counts.
On an instance, \MRG{} cannot distinguish two unmatched vertices of the same class and local state, so the whole run is described by a few class-level counts.

\paragraph{Step 2: Derive and justify a differential equation:}
In the continuous time formulation of \cref{sec:activation}, every active vertex acts at rate $1$, so a class with $hn$ active vertices contributes actions at rate $hn$, without a denominator involving the total number of active vertices.
After dividing the counts by $n$, these rates give a candidate system of ODEs.
Wormald's theorem (\cref{thm:wormald}), applied to the discrete process with the rates converted back to expected changes per step, shows that with high probability the normalized counts stay close to its solution, called their \emph{fluid limit}.
We demonstrate the complete argument on the classical bomb graph in \cref{sec:bomb-proof}.

\paragraph{Step 3: Count the remaining matching edges:}
The drift may stop being Lipschitz when a class is exhausted, so the ODE approximation alone does not necessarily determine the final matching.
We stop the approximation while the relevant counts are still at least $\eps n$, and bound the difference between the remaining contribution and an explicit function of the counts at that time.
For the Double-Bomb graphs, maximality of the output matching is enough for this step.
For the paw with a pool, we also use the independent uniformly random relative orders within the remaining disjoint paws.
Letting $n\to\infty$ and then $\eps\to 0$ gives the limiting approximation ratio.

We collect these steps, namely the count chain, its drift, the ODE trajectory, and the end phase, in the certification lemma (\cref{lem:template}), which proves all four of our main results.
For the Double-Bomb families, the ODE admits a closed form and the limiting ratio is an explicit one-dimensional integral.

\paragraph{Scope of the method:}
Wormald's theorem and the continuous time reformulation are established tools.
Our contribution is their use with counts that retain the sparse gadget edges of a structured worst-case graph, together with a separate analysis of the end phase, in a form that is reusable when the graph or the greedy rule changes.

\subsection{Overview of the Double-Bomb Analysis}
\label{sec:glance}

We now illustrate our toolkit on $S_n(1)$.
The construction of $S_n(\sigma)$ is due to Tang, Wu, and Zhang~\cite{TWZ23}, and our observation is that it also gives an upper bound below $2/3$ for \MRG{}.
Here we derive the candidate limit and explain the contribution from the end phase.
The full proof is given in \cref{sec:mrg}.

\paragraph{The rates of change:}
Use the continuous time formulation described above, and divide all counts by $n$.
Write $h(s)$ for the candidate limiting density of active hubs on each side, and $q(s)$ for that of active satellites on each side.
The symmetry here is a property of the deterministic ODE and the graph structure, rather than an assertion that the random counts on the two sides are equal.
Initially, $h(0)=q(0)=1$.

While $h>0$, a fixed active hub $x_i\in X$ can be matched in four ways: it acts itself (rate $1$), its antenna acts (rate $1$), a hub of $Y$ selects it (limiting rate $h/(h+q)$), or a satellite of $P$ selects it (limiting rate $q/h$).
The last possibility removes a hub before its antenna can use it.
Adding these contributions gives
\begin{equation}
h'=-h\left(2+\frac{h}{h+q}+\frac{q}{h}\right)
=-2h-\frac{h^2}{h+q}-q.
\label{eq:primer-ODE-x}
\end{equation}

An active satellite disappears when it acts (rate $1$), or when a hub selects it (limiting rate $h/(h+q)$).
Its partner chooses it with probability at most $1/(hn+1)$ when that partner acts, so this contribution vanishes on intervals where $h$ is bounded away from zero.
Thus,
\begin{equation}
q'=-q\left(1+\frac{h}{h+q}\right).
\label{eq:primer-ODE-q}
\end{equation}

\paragraph{Solving the ODE:}
Let $r=q/h$.
Since $r'/r = q'/q - h'/h$, \cref{eq:primer-ODE-x,eq:primer-ODE-q} together give $r'/r=1+r$.
Since $r(0)=1$, we obtain $r=e^s/(2-e^s)$.
Substituting $v=e^s$ and $h=q(2-v)/v$ into \cref{eq:primer-ODE-q} gives,
\begin{equation*}
\begin{aligned}
	v\frac{dq}{dv}=-\frac{q}{2}(4-v), \text{ implying } q&=v^{-2}e^{(v-1)/2}, \text{ and}\\
	h&=v^{-3}(2-v)e^{(v-1)/2}.
\end{aligned}
\end{equation*}

The hub density therefore first reaches zero at $s^\ast=\log 2$, while the satellite density tends to $q(s^\ast)=\sqrt e/4>0$.
This is precisely where a separate end-phase argument is needed.

\paragraph{Counting the matching edges:}
Before time $s=s^\ast$, edges are created at normalized rate $4h+2q$, because an active hub, its antenna, or an active satellite creates an edge when it acts.
To determine the remaining contribution, we also track intact satellite pairs.
If $b(s)$ is their candidate limiting density, its rate equation is $b'=-2(1+h/(h+q))b$, with $b(0)=1$.
Comparison with \cref{eq:primer-ODE-q} gives $b=q^2$.

After the hubs are removed, the remaining non-isolated vertices form disjoint satellite edges, all of which a maximal matching must take.
The limiting contribution from this phase is therefore $b(s^\ast)=e/16$.
In the proof, we stop with at most $O(\eps n)$ hubs and show that their subsequent actions change this count by at most $O(\eps n)$.
Since $\nu(S_n(1))=3n$, integrating $4h+2q$ and adding the terminal contribution gives
\begin{equation}
\begin{aligned}
	\rho_S(1)&=\frac13\Big(\int_1^2\Big(\frac8{v^4}-\frac2{v^3}\Big)e^{(v-1)/2}\, d v+\frac e{16}\Big)\\
	&=0.66298\ldots<\frac23.
\end{aligned}
\label{eq:primer-rho_S-1}
\end{equation}

This already shows how the Double-Bomb can improve on the classical $2/3$ bound.
Allowing the satellite density $\sigma$ to vary gives the stronger bound in \cref{thm:mrg-bip}, and adding cliques on the hubs gives \cref{thm:mrg-general}.

\subsection{Certification Lemma}
\label{sec:template}

The double-bomb graph calculation in \glanceref{} illustrates the two parts of the argument.
Wormald's theorem controls the random trajectory while the relevant classes have positive density.
Once those classes become small, a direct count controls what remains.
The following lemma keeps these two requirements separate.

Let $\mathcal{A}\in\{\MRG,\RDO\}$ run in step form on $G_n$, and let $M_t$ be the matching after $t$ steps.
A \emph{count vector} is a process $Z_t\in\mathbb{Z}_{\ge 0}^a$ whose last coordinate is $|M_t|$ and from which the number $|U_t|$ of active vertices is a fixed linear function, $|U_t|=n\,u(Z_t/n)$.
We write $z=Z/n$ for the rescaled counts and use lower-case letters for its coordinates, for instance for the coordinate $H_X$ of $Z_t$ we write $h_X=H_X/n$.
The last coordinate of $z$ is denoted by $\mu=|M_t|/n$.
Throughout, $|V(G_n)|=O(n)$, so every coordinate of $Z_t$ is $O(n)$ and every run has $O(n)$ steps.
After the algorithm finishes we set $Z_{t+1}=Z_t$.

\begin{lemma}[Certification lemma]
\label{lem:template}
Let $Z_t$ be a count vector with $Z_0=nz_0+O(1)$, and let $h$ be a linear function of $z$ (the \emph{budget}).
Suppose that there are constants $K,\beta>0$, independent of $\eps$, and for every small $\eps>0$ a bounded open box $D_\eps\subset\mathbb{R}^a$ containing $z_0$, a map $g\colon D_\eps\to\mathbb{R}^a$, and constants $c_\eps,C_\eps>0$, such that the following hold.
\begin{enumerate}
	\item[(C1)](Counts and bounded jumps) While $Z_t/n\in D_\eps$, the conditional law of $Z_{t+1}-Z_t$ given the history is a function of $Z_t$, and $\|Z_{t+1}-Z_t\|_\infty\le\beta$.
	\item[(C2)] (Drift and regularity) While $Z_t/n\in D_\eps$,
	\[
	\Big\|\E[Z_{t+1}-Z_t\mid\mathcal{F}_t]-\frac{g(Z_t/n)}{u(Z_t/n)}\Big\|_\infty\le\frac{C_\eps}{n}.
	\]
	Moreover $g$ is Lipschitz and $u\ge c_\eps$ on $D_\eps$.
	\item[(C3)] (Interior trajectory) The solution $\bar z$ of 
	\[\frac{d\bar z}{ds}=g(\bar z), \text{ with }\bar z(0)=z_0\]
	first reaches the budget $\eps$ at a finite time $s_\eps$, that is, $h(\bar z(s))>\eps$ for $s<s_\eps$ and $h(\bar z(s_\eps))=\eps$, and $\bar z(s)\in D_\eps$ for all $s\in[0,s_\eps]$.
	\item[(C4)] (End phase) There is a Lipschitz function $\Phi$ such that, for every sequence of steps $t_1=t_1(n)$, with probability $1-o(1)$ the following holds: if $\|Z_{t_1}/n-\bar z(s_\eps)\|_\infty\le\eps$, then
	\[
	\big|\,|M|-|M_{t_1}|-n\,\Phi(Z_{t_1}/n)\big|\le K\eps n+o(n).
	\]
\end{enumerate}
Let $\bar\mu$ be the last coordinate of $\bar z$.
If $\Lambda=\lim_{\eps\to 0}\big(\bar\mu(s_\eps)+\Phi(\bar z(s_\eps))\big)$ exists, then $|M|/n\to\Lambda$ in probability and $\E|M|/n\to\Lambda$.
\end{lemma}

Conditions (C1) and (C2) supply the hypotheses of Wormald's theorem.
The bounded-jump part of (C1) is (W1), as the Markov property lets us compute the drift from the counts alone.
For \MRG{} on an instance of a template, this follows from interchangeability of vertices in the same class and local state.
For \RDO{} with fixed preferences it must be checked separately (see \cref{rem:tiers}).
Condition (C2) supplies both parts of (W2), which are the trend estimate and Lipschitz regularity of $f=g/u$.
The positivity of $u$ ensures that dividing by the number of active vertices does not introduce a singularity on $D_\eps$.

The map $g$ records rates in log-time, whereas $g/u$ records expected changes per discrete step.
Indeed, the active vertices act at total rate $|U|=nu$, so the conditional mean waiting time between steps is $1/(nu)$.
Dividing the rate $ng$ of change of the counts by this total action rate gives $g/u$.
We write $g_{h_X}$ for the $h_X$-coordinate of $g$, and similarly for other coordinates.
Lipschitz continuity can be checked one coordinate at a time.

Conditions (C3) and (C4) are additional requirements for determining the final output.
Condition (C3) identifies the ODE trajectory and an interior point at which to stop the approximation; local existence and uniqueness already follow from the Lipschitz assumption in (C2).
In the applications, the domain has threshold $\eps/2$, so the trajectory up to budget $\eps$ remains strictly inside it.
Condition (C4) then estimates the matching edges added after that point.
It is a separate counting statement, rather than another hypothesis of Wormald's theorem.

\begin{proof}
Fix $\eps>0$. Let $f:=g/u$, which is Lipschitz on $D_\eps$ because $g$ is Lipschitz and bounded there and $u\ge c_\eps$.
Let $\tau(s)=\int_0^s u(\bar z(r)) dr$.
Since $u>0$, the map $s\mapsto\tau(s)$ is increasing, and by the chain rule $z(\tau(s))=\bar z(s)$ solves $dz/d\tau=f(z)$.
Take the domain $D=(-1,L)\times D_\eps$, where the constant $L$ exceeds $\tau(s_\eps)$ and the $O(1)$ bound on the number of steps divided by $n$.
By (C3), the compact path $\{(\tau(s),\bar z(s)):0\le s\le s_\eps\}$ lies in the open set $D$, so it has a positive distance $d_\eps$ from $\partial D$.
Apply \cref{thm:wormald} with $\lambda^\ast=C_\eps/n$ and $\lambda=n^{-1/4}$. 
The failure probability is $O(n^{1/4}e^{-n^{1/4}/\beta^3})=o(1)$.
(The initial condition $Z_0/n=z_0+O(1/n)$ changes the solution by $O(1/n)$, since $f$ is Lipschitz.)
For large $n$ we have $C\lambda<d_\eps$, hence $\tau_\lambda\ge\tau(s_\eps)$.
Let $t_1=\lfloor n\,\tau(s_\eps)\rfloor$, a deterministic step with $t_1\le\tau_\lambda n$.
With probability $1-o(1)$ we get $Z_{t_1}=n\,z(t_1/n)+O(\lambda n)=n\bar z(s_\eps)+o(n)$, and in particular $\|Z_{t_1}/n-\bar z(s_\eps)\|_\infty\le\eps$ for large $n$.
By (C4) and the Lipschitz property of $\Phi$,
\begin{equation}
	\Big|\frac{|M|}{n}-\bar\mu(s_\eps)-\Phi(\bar z(s_\eps))\Big|\le K\eps+o(1)
\end{equation}
with probability $1-o(1)$.
Letting $\eps\to 0$ gives convergence in probability, and since $|M|/n$ is bounded, also convergence of the expectation.
This completes the Proof of Lemma~\ref{lem:template}.
\end{proof}

\subsection{The Bomb Graph via Lemma~\ref{lem:template}}
\label{sec:bomb-proof}
We now prove the classical bomb graph bound~\cite{DF91} with the certification lemma to demonstrate the method.

\begin{proposition}
\label{prop:bomb}
$\E|M_{\MRG}(B_n)|/\nu(B_n)\to 2/3$ as $n\to\infty$.
\end{proposition}

\begin{proof}
We check the four conditions of \cref{lem:template}.

\paragraph{(C1) The chain:}
Let $Z=(H_X,H_Y,W_X,W_Y,|M_t|)$, where $H_X,H_Y$ are the numbers of active hubs $|X\cap U|, |Y\cap U|$, and $W_X,W_Y$ the numbers of active antennas $|X'\cap U|, |Y'\cap U|$.
The $n$-rescaled counts are $z=(h_X,h_Y,w_X,w_Y,\mu)$.
Every active hub has an active antenna, and an active antenna is either attached to an active hub or isolated.
Let us look at the possible steps on the $X$ side.
\begin{itemize}
	\item A hub of $X$ acts (probability $H_X/U$). It takes its antenna with probability $1/(H_Y+1)$, and a hub of $Y$ otherwise.
	\item An antenna of $X$ whose hub is active acts (probability $H_X/U$) and takes its hub.
	\item An isolated antenna of $X$ acts (probability $(W_X-H_X)/U$).
\end{itemize}
Here $U=H_X+H_Y+W_X+W_Y$.
These probabilities depend only on $Z$, and every coordinate changes by at most $2$.
The same argument can be symmetrically repeated for the $H_Y$ and $W_Y$ pair.

\paragraph{(C2) The drift:}
Let the budget be $h=h_X$, and let $D_\eps$ be the box
\begin{equation}
	\label{eq:B_n-domaindef}
	\begin{aligned}
		D_\eps := \{z=(h_X,h_Y,&w_X,w_Y,\mu): h_X,h_Y\in(\eps/2,2);\\ &w_X,w_Y\in(-\eps/8,2); \mu\in(-1,3)\}
	\end{aligned}
\end{equation}
The exact expected change of $H_X$ in one step is
\begin{equation}
	\E[H_X(t+1)-H_X(t)\mid\mathcal{F}_t]=-\frac1U\Big(2H_X+H_Y\cdot\frac{H_X}{H_X+1}\Big),  
\end{equation}
since $H_X$ decreases by one when a hub of $X$ or its antenna acts, and when a hub of $Y$ acts and takes a hub of $X$ rather than its own antenna.
Replacing $H_X/(H_X+1)$ by $1$, that is, dropping the antenna option of an acting hub, changes this by at most $1/(H_X+1)\le 2/(\eps n)$ on $D_\eps$, and the same holds for the other coordinates.
So (C2) holds with $C_\eps=2/\eps$ and
\begin{equation}
	g(z)=\big(-2h_X-h_Y,\ -2h_Y-h_X,\ -w_X,\ -w_Y,\ 2h_X+2h_Y\big),  
\end{equation}
listing the coordinates $g_{h_X},g_{h_Y},g_{w_X},g_{w_Y},g_\mu$ in this order, with $u=h_X+h_Y+w_X+w_Y$.
In $g_{h_X}$, the term $-2h_X$ accounts for the actions of $X$-hubs and of their antennas, and $-h_Y$ for the $Y$-hubs that take an $X$-hub with them.
The map $g$ is linear, so it is Lipschitz, and $u>\eps-\eps/4>\eps/2$ on $D_\eps$.

\paragraph{(C3) The solution:}
The ODE and $z_0=(1,1,1,1,0)$ are symmetric under $X\leftrightarrow Y$, so by uniqueness $h_X=h_Y=h$ with $h'=-3h$.
Thus $h=e^{-3s}$, $w_X=w_Y=e^{-s}$ and $\bar\mu(s)=\frac43(1-e^{-3s})$.
The budget first reaches $\eps$ at $s_\eps=\frac13\log(1/\eps)$. Up to that time $h_X=h_Y\in[\eps,1]$, $w_X,w_Y\in[0,1]$ and $\bar\mu\in[0,\frac43]$, so the path stays strictly inside $D_\eps$.

\paragraph{(C4) The end phase:}
If $Z_{t_1}/n$ is $\eps$-close to $\bar z(s_\eps)$, then $H_X,H_Y\le 2\eps n$. After step $t_1$, every new edge of $M$ contains a hub that is active at $t_1$, so $0\le|M|-|M_{t_1}|\le H_X+H_Y\le 4\eps n$.
So $\Phi=0$.

The certification lemma gives $|M|/n\to\lim_{\eps\to 0}\frac43(1-\eps)=\frac43$, and $\nu(B_n)=2n$ implies that \MRG{} is at most $2/3$-approximate.
\end{proof}

\subsection{Computational Experiments}
\label{sec:toolkit-search}

We used numerical integration to explore templates and to check candidate formulas before proving them, tracking the mass of gadget copies in each local state under \MRG{}, or under class-level preference tiers for \RDO{}.
For a general template, a class becoming exhausted may change the form of the ODE, so this evaluator is heuristic and not by itself a proof.

The experiments helped identify the clique modification of the Double-Bomb and the paw with a pool, starting from the constructions of Tang, Wu, and Zhang~\cite{TWZ23}.
We checked candidates with finer integration steps, simulations of the count chain, and runs on explicit graphs; the bounds proved here follow from \cref{lem:template} and the subsequent analysis.
The implementation and data are in the repository~\cite{Mukherjee_Fluid_limits_2026}, and further numerical evidence is discussed in \cref{sec:conclusion} and drawn in \cref{fig:candidates}.

\section{Random Decision Order}
\label{sec:rdo}

We apply \cref{lem:template} to prove \cref{thm:rdo-bip,thm:rdo-general}.
We first explain how fixed preferences affect the count chain, then analyze $S_n(\sigma)$ and $PP_n(\beta)$.

\subsection{Preferences and the Count Chain}
\label{sec:rdo-prefs}

In \RDO, every vertex has a fixed preference order.
In our search we also allowed class-level preferences, where each class ranks its options in \emph{tiers}, and an acting vertex chooses uniformly among the active neighbors in its first nonempty tier.
The following remark explains how this relates to fixed preferences, and why the count-chain condition needs to be checked additionally by hand.

\begin{remark}[Class-level versus fixed preferences]
\label{rem:tiers}
Uniform choice within a tier can be realized by giving each vertex an independent uniformly random order within each of its tiers, drawn independently of the decision order.
Until a vertex acts, its own preference order has not been used, so the active set does not depend on that order.
Revealing it at that point gives a uniformly random active neighbor in the first nonempty tier.
If $\Pi$ denotes the random preference profile and $\E_\Pi\E_{\mathrm{ord}}|M(G,\Pi)|\le\alpha\nu(G)$, then some fixed profile $\pi$ satisfies $\E_{\mathrm{ord}}|M(G,\pi)|\le\alpha\nu(G)$, where $\E_{\mathrm{ord}}$ is over the decision order.
And so, a proved expectation bound for the tiered process gives an upper bound for \RDO{} with fixed preferences.

For a fixed profile, however, the class counts need not determine the conditional law of the next step.
For example, take vertices $u,v,p_1,q_1,p_2,q_2$, with edges $p_iq_i$, $uq_i$ and $vp_i$ for $i\in\{1,2\}$, and the common preference order
\[
v\succ p_1\succ p_2\succ q_1\succ q_2\succ u.
\]
Let us track whether $u,v$ are active, the matching size, and the numbers of pairs $p_iq_i$ with both endpoints active, only $p_i$ active, or only $q_i$ active.
If $p_2$ acts first, it takes $v$, leaving pair~$1$ intact and only $q_2$ active in pair~$2$.
If $p_1$ acts first, the roles of the pairs are exchanged, and all these counts are the same.
In the first state, an action of $u$ takes $q_1$ and creates a pair with only its $p$-endpoint active.
This happens at the next step with probability $1/4$, whereas in the second state no next step creates such a pair.
So the counts do not satisfy (C1) of \cref{lem:template}.

In both of our \RDO{} theorems, every choice made before the end phase is among vertices that are interchangeable for the counts.
We verify this directly in each proof.
\end{remark}

\subsection{The Double-Bomb}
\label{def:sat}

Recall the Double-Bomb $S_n(\sigma)$ from the introduction (\cref{fig:constructions}(b)) and~\cite[Appendix~B.1]{TWZ23}.
For now take $\sigma>0$, and let $m=\lfloor\sigma n\rfloor$.
It has hubs $X=\{x_i\}$ and $Y=\{y_i\}$ and antennas $X'=\{x'_i\}$ and $Y'=\{y'_i\}$ for $i\in[n]$, satellites $P=\{p_j\}$ and $Q=\{q_j\}$ for $j\in[m]$, the perfect matching $M^*=\{x_ix'_i\}\cup\{y_iy'_i\}\cup\{p_jq_j\}$, and all edges between $X$ and $Y$, between $X$ and $P$, and between $Y$ and $Q$.
The Double-Bomb with cliques $S^K_n(\sigma)$ (\cref{fig:constructions}(c)) additionally contains all edges inside $X$ and all edges inside $Y$.

So $\nu=2n+m=(2+\sigma)n+O(1)$ for both graphs, and $S_n(\sigma)$ is bipartite, with sides $X\cup Q\cup Y'$ and $Y\cup P\cup X'$.
We analyze $S_n(\sigma)$ with the \emph{hub-first} preferences of~\cite{TWZ23}:
\begin{itemize}
\item A hub prefers the hubs of the other side, then its satellites, then its antenna (for example, $Y\succ_x P\succ_x X'$ for every $x\in X$).
\item A satellite prefers its hubs, then its partner (for example, $X\succ_p Q$ for every $p\in P$).
\end{itemize}
Ties inside a class are broken by index.
The authors ran this graph with up to $1000$ satellite pairs.
Among the parameters they tested, the smallest ratio, about $0.646$, occurred when the number of hubs per side was $1.5$ times the number of satellite pairs, corresponding to $\sigma=2/3$.

\paragraph{Intuition:}
We now give an overview of the upper bound proof.
Let us redo the computation of \glanceref{} for the hub-first preferences.
While both sides have active hubs, a hub of $X$ takes a hub of $Y$, and similarly for $Y$.
A satellite is therefore matched only when it acts itself, and then it takes a hub of its own side.
Let $N_{11}$ count satellite pairs $p_jq_j$ with both endpoints active, and let $N_{10}$ and $N_{01}$ count pairs with only $p_j$ or only $q_j$ active, respectively.
Thus $P$ has $N_{11}+N_{10}$ active satellites.

Since ties inside a class are broken by index, all satellites of $P$ choose the first active hub of $X$.
So the uniform per-hub rate $q/h$ used for \MRG{} in \glanceref{} does not apply.
We can still count all actions that remove a hub of $X$.
Such a hub disappears whenever a hub of $X$ or its antenna acts ($2H_X$ possible actors), a hub of $Y$ acts ($H_Y$ possible actors), or a satellite of $P$ acts ($N_{11}+N_{10}$ possible actors).

The candidate fluid trajectory is symmetric between the two sides.
Writing $h$ for the hub density on each side and $q$ for the satellite density on each side gives $h'=-3h-q$ and $q'=-q$ in log-time.
In particular, $q=\sigma e^{-s}$, and the satellites remove hubs at a total rate that does not depend on how many hubs remain.
The solution reaches $h=0$ at a finite time $s^\ast$, with $q(s^\ast)>0$, so fewer antennas get matched.

After all hubs are removed, the remaining non-isolated vertices form disjoint satellite edges.
So, we only count the intact satellite pairs at $s^\ast$ as the terminal contribution.
In the finite process the two hub sides do not necessarily run out at the same time, so we stop earlier at the fluid cutoff $h=\eps$, and bound the effect of the remaining hubs in (C4).

\subsection{Proof of \texorpdfstring{\cref{thm:rdo-bip}}{the bipartite bound}}

We check the four conditions of \cref{lem:template}.

\paragraph{(C1) The chain:}
Let
\[
Z=(H_X,H_Y,W_X,W_Y,N_{11},N_{10},N_{01},|M_t|),
\]
where
\begin{itemize}
\item $H_X,H_Y$ count active hubs,
\item $W_X,W_Y$ count active antennas,
\item $N_{11},N_{10},N_{01}$ are the active satellite-pair counts defined above, and
\item $|M_t|$ is the number of edges of the current matching.
\end{itemize}
Every active hub has an active antenna.
While both $H_X$ and $H_Y$ are positive, the steps with an actor on the $X$ side are the following.
\begin{itemize}
\item A hub of $X$ acts ($H_X$ choices) and takes a hub of $Y$, so $H_X,H_Y$ decrease by $1$.
\item An antenna of $X$ with an active hub acts ($H_X$ choices) and takes its hub, so $H_X,W_X$ decrease by $1$.
\item An isolated antenna of $X$ acts ($W_X-H_X$ choices) and leaves, so $W_X$ decreases by $1$.
\item A satellite $p_j$ of an intact pair acts ($N_{11}$ choices) and takes a hub of $X$, so $H_X,N_{11}$ decrease and $N_{01}$ increases by $1$.
\item A satellite $p_j$ whose partner is matched acts ($N_{10}$ choices) and takes a hub of $X$, so $H_X,N_{10}$ decrease by $1$.
\end{itemize}
Every step with a partner creates an edge, and the $Y$ side is the mirror image under $X\leftrightarrow Y$, $P\leftrightarrow Q$, $N_{10}\leftrightarrow N_{01}$.
Each actor is chosen with probability $1/U$, where $U=H_X+H_Y+W_X+W_Y+2N_{11}+N_{10}+N_{01}$.
Which vertex of a class is chosen as the partner does not change the counts, because all active hubs of a side are alike (each has an active antenna).
So the law of the next step depends only on $Z$, even though the preferences inside a class are a fixed order.
Every coordinate changes by at most $1$, so (C1) holds.
The count closure uses these particular preferences (\cref{rem:tiers}).

\paragraph{(C2) The drift:}
Let $B=5+2\sigma$ and let $D_\eps$ be the box where $h_X,h_Y\in(\eps/2,B)$ and $w_X,w_Y,n_{11},n_{10},n_{01},\mu\in(-\eps/8,B)$. On $D_\eps$ both hub sides are nonempty, the steps above are exact, and the log-time rates are
\begin{align*}
g_{h_X}&=-2h_X-h_Y-(n_{11}+n_{10}), & g_{w_X}&=-w_X,\\
g_{n_{11}}&=-2n_{11}, & g_{n_{01}}&=n_{11}-n_{01},\\
g_\mu&=2h_X+2h_Y+2n_{11}+n_{10}+n_{01},
\end{align*}
and, by the mirror symmetry, $g_{h_Y}=-2h_Y-h_X-(n_{11}+n_{01})$, $g_{w_Y}=-w_Y$ and $g_{n_{10}}=n_{11}-n_{10}$.
The conditional expected increment of the count vector $Z_t$ in one discrete step is $\E[Z_{t+1}-Z_t\mid\mathcal{F}_t]=g(Z_t/n)/u(Z_t/n)$.
Thus the trend error in (C2) is zero.
The coordinates of $g$ are linear, so $g$ is Lipschitz.
Also $u=h_X+h_Y+w_X+w_Y+2n_{11}+n_{10}+n_{01}>\eps-6\eps/8=\eps/4$ on $D_\eps$.

\paragraph{(C3) The solution:}
The ODE and $z_0=(1,1,1,1,\sigma,0,0,0)$ are symmetric, so by uniqueness $h_X=h_Y=h$ and $n_{10}=n_{01}$.
Let $q=n_{11}+n_{10}$ be the mass of active satellites of $P$, which by symmetry equals that of $Q$.
Then
\[
h'=-3h-q,\, q'=-q\, \text{ and } n_{11}'=-2n_{11},
\]
so that $q=\sigma e^{-s}$, $n_{11}=\sigma e^{-2s}=q^2/\sigma$, and we have
\[
h(s)=\Big(1+\frac\sigma2\Big)e^{-3s}-\frac\sigma2e^{-s}.
\]
The hub density first reaches zero at time
\[
s^\ast=\frac12\log\frac{2+\sigma}{\sigma}, \text{ where }w=e^{-s^\ast}=\sqrt{\frac{\sigma}{2+\sigma}}.
\]
The edge rate is $g_\mu=4h+2q$.
For the budget $h=h_X$, the solution first reaches $h=\eps$ at a time $s_\eps\to s^\ast$, since $h'<0$ before $s^\ast$.
Up to then $h_Y=h_X\in[\eps,1]$, $w_X=w_Y=e^{-s}\in[0,1]$ and $n_{11},n_{10},n_{01}\in[0,\sigma]$.
Moreover, $\bar\mu'=4h+2q\le-2h'$, so $\bar\mu(s)\le 2(1-h(s))\le 2$.
So the path stays strictly inside $D_\eps$.

\paragraph{(C4) The end phase:}
Let $t_1$ be a step at which $Z_{t_1}/n$ is $\eps$-close to $\bar z(s_\eps)$, so that $H_X,H_Y\le 2\eps n$.
Every edge of $G_n$ joins a hub to a hub, antenna or satellite, or is a satellite edge $p_jq_j$.
So every edge added after $t_1$ either contains a hub active at $t_1$, and there are at most $H_X+H_Y$ such edges, or is the edge $p_jq_j$ of a pair counted in $N_{11}(t_1)$.
Conversely, if an intact pair at $t_1$ does not become an edge of $M$, maximality implies that at least one of its endpoints is matched elsewhere.
Its only other neighbors are hubs, so at most $H_X+H_Y$ intact pairs can be lost this way.
Therefore, deterministically,
\[
N_{11}(t_1)-4\eps n\le|M|-|M_{t_1}|\le N_{11}(t_1)+4\eps n,
\]
and (C4) holds with $\Phi=n_{11}$.

\paragraph{The value:}
Since $4h+2q=(4+2\sigma)e^{-3s}$, the certification lemma (Lemma~\ref{lem:template}) gives
\[
\begin{aligned}
\Lambda&=\int_0^{s^\ast}(4h+2q)\, d s+n_{11}(s^\ast)\\
&=\frac23(2+\sigma)(1-w^3)+\sigma w^2.
\end{aligned}
\]
By $\nu/n\to 2+\sigma$ and $\sigma/(2+\sigma)=w^2$, this gives us
\[
\rho_{\RDO}(\sigma)=\frac23-\frac23w^3+w^4.
\]
Its derivative in $w$ is $2w^2(2w-1)$, which is negative for $0<w<1/2$ and positive for $1/2<w<1$.
Since $w$ increases with $\sigma$, the minimum occurs at $w=1/2$, or $\sigma=2/3$.
The minimum value is $\frac23-\frac1{12}+\frac1{16}=\frac{31}{48}$, as desired.
\qed

\begin{remark}
\label{rem:rdo-bip-search}
Note that $\sigma=0$, the satellite counts vanish and $h=e^{-3s}$.
The same argument applies with $s_\eps=\frac13\log(1/\eps)$ and $\Phi=0$, giving $\Lambda=4/3$ and ratio $2/3$.
Thus the formula extends to $\sigma=0$, and $\rho_{\RDO}(\sigma)<2/3$ exactly when $0<w<2/3$, or $0<\sigma<8/5$.
Additionally, we note that computation suggests that the upper bound for \RDO{} on bipartite graphs could be lowered further (see \cref{sec:conclusion} and \cref{fig:candidates}(c)).
\end{remark}

\subsection{General Graphs and the Paw with a Pool}
\label{sec:pawpool}

Tang, Wu, and Zhang~\cite[Theorem~B.1]{TWZ23} take the paw, the triangle $abc$ plus the edge $cd$, with the common preference order $c\succ b\succ a\succ d$.
Unless $d$ acts first among the four vertices, the first vertex to act takes $c$, or is $c$ and takes $b$, and only one edge is formed.
Thus the expected matching size is $1\cdot\frac34+2\cdot\frac14=\frac54$, against $\nu=2$, giving the ratio $5/8$.

\paragraph{The pool:}
The weak point of this example, from the adversary's point of view, is the probability $1/4$ that $d$ acts first.
We reduce this advantage by giving $d$ a more attractive option.
Fix $\beta>0$, take $n$ disjoint paws $a_ib_ic_id_i$, and add a \emph{pool} of $\lfloor\beta n\rfloor$ disjoint edges $e_jf_j$.
Make $E=\{e_j\}$ a clique, and join every $d_i$ to every $e_j$, as in \cref{fig:constructions}(d).
All vertices use the common order
\[
E\succ C\succ B\succ A\succ F\succ D,
\]
with ties inside a class broken by index.
In particular, $d_i$ prefers the pool to $c_i$, and $e_j$ prefers the other pool vertices to $f_j$.
The perfect matching $\{a_ib_i,c_id_i,e_jf_j\}$ shows $\nu=2n+\lfloor\beta n\rfloor$.

If $d_i$ acts first in its paw while a pool vertex is active, it takes a pool vertex and leaves a triangle, so that paw contributes only one internal edge.
Also, while at least two pool vertices are active, an acting $e_j$ takes another pool vertex and strands both pendants.
The pool size $\beta$ controls how long these effects persist, and also changes the size of the maximum matching.

\begin{proposition}
\label{prop:pawpool}
Let $w=(1+2\beta)^{-1/2}$.
For the graphs above, as $n\to\infty$,
\begin{equation}
	\frac{\E|M_{\RDO}|}{\nu}\to F(w)=\frac{2w^2+\tfrac12w^6+\tfrac23(1-w^3)}{3w^2+1}.
	\label{eq:PP-RDO-F(w)}
\end{equation}
\end{proposition}

\begin{proof}
As usual, we check the four conditions of \cref{lem:template}.

\paragraph{(C1) The chain:}
Call a paw \emph{intact} if its four vertices are active, and a \emph{triangle} if exactly $a,b,c$ are active.
Let $N_I$ and $N_T$ count intact paws and triangles, and let $N_2$ count paws whose only active vertices are $a,b$.
The last state can arise when $d$ takes $c$ after the pool empties, and we include it to keep the count vector valid throughout the run.
Let $D$ count active vertices $d_i$ whose $c_i$ is matched, and let $Y$ count active pool vertices $e_j$.
Each active $e_j$ still has its pendant $f_j$ active.
Let $W$ count all other active isolated vertices, excluding those already counted in $D$.
Our count vector is
\[
Z=(N_I,N_T,N_2,D,Y,W,|M_t|).
\]
We write $\omega=W/n$, reserving $w$ for the parameter in the proposition.
While $Y\ge 2$, the possible steps are as follows.
\begin{itemize}
	\item One of $a,b,c$ of an intact paw acts ($3N_I$ choices).
	An acting $a$ or $b$ takes $c$.
	An acting $c$ takes $b$, since $B\succ A\succ D$ in the common preference order.
	In each case, the remaining vertex of $\{a,b\}$ becomes isolated.
	Thus $N_I$ decreases by $1$, and $D,W$ increase by $1$.
	\item The vertex $d$ of an intact paw acts ($N_I$ choices) and takes a pool vertex, stranding its pendant.
	Thus $N_I,Y$ decrease by $1$, and $N_T,W$ increase by $1$.
	\item A vertex of a triangle acts ($3N_T$ choices) and matches inside it, so $N_T$ decreases and $W$ increases by $1$.
	\item A vertex counted in $D$ acts ($D$ choices) and takes a pool vertex, so $D,Y$ decrease and $W$ increases by $1$.
	\item A pool vertex acts ($Y$ choices) and takes another pool vertex, so $Y$ decreases and $W$ increases by $2$.
	\item A pendant $f_j$ with $e_j$ active acts ($Y$ choices) and takes $e_j$, so $Y$ decreases by $1$.
	\item A vertex counted in $W$ acts ($W$ choices) and leaves, so $W$ decreases by $1$.
\end{itemize}
All steps except the last create one edge.
No vertex outside a paw takes a paw vertex, since a pool vertex prefers another pool vertex, or its own pendant, to any $d_i$.
In particular, $N_2=0$ until the pool empties.
Which pool vertex is taken does not change the counts, so the conditional law of the next step depends only on $Z$.
Every coordinate changes by at most $2$.
Each actor has probability $1/U$, where
\[
U=4N_I+3N_T+2N_2+D+2Y+W.
\]

\paragraph{(C2) The drift:}
For $0<\eps<\beta$, let $B=5+2\beta$, and let $D_\eps$ be the box where $y\in(\eps/2,B)$ and $n_I,n_T,n_2,d,\omega,\mu\in(-\eps/16,B)$.
For large $n$, every reachable state in $D_\eps$ has $Y>\eps n/2\ge 2$, so the steps above are exact.
The log-time rates are
\begin{equation}
	\begin{aligned}
		g_{n_I}&=-4n_I, & g_{n_T}&=n_I-3n_T,\\
		g_{n_2}&=0, & g_d&=3n_I-d,\\
		g_y&=-3y-(n_I+d), &g_\mu&=4n_I+3n_T+d+2y,
	\end{aligned}
	\label{eq:PP-RDO-LogRates}
\end{equation}
and $g_\omega=4n_I+3n_T+d+2y-\omega$.
In $g_y$, the term $-2y$ comes from acting pool vertices, $-y$ from acting pendants, and $-(n_I+d)$ from acting $d$'s.
The conditional expected increment of the count vector $Z_t$ in one discrete step is $\E[Z_{t+1}-Z_t\mid\mathcal{F}_t]=g(Z_t/n)/u(Z_t/n)$. Thus the trend error in (C2) is zero.
The map $g$ is linear and hence Lipschitz, and on $D_\eps$ we have
\[
u=4n_I+3n_T+2n_2+d+2y+\omega>\eps-\frac{11\eps}{16}=\frac{5\eps}{16}.
\]
This verifies (C2).

\paragraph{(C3) The solution:}
The initial condition is $z_0=(1,0,0,0,\beta,0,0)$.
From \cref{eq:PP-RDO-LogRates} we get $n_I=e^{-4s}$, $n_T=e^{-3s}-e^{-4s}$ and $n_2=0$.
Also $n_I+d=e^{-s}$, since an active $d$ is removed only by its own action while the pool is nonempty.
Then $y'=-3y-e^{-s}$ gives
\[
y(s)=\Big(\beta+\frac12\Big)e^{-3s}-\frac12e^{-s},
\]
which vanishes at the time $s^\ast$ such that $e^{-2s^\ast}=1/(1+2\beta)$, or $e^{-s^\ast}=w$.
The budget is $h=y$.
The walk first reaches $\eps$ at a time $s_\eps\to s^\ast$, since $y'=-3y-e^{-s}<0$ while $y>0$.
The remaining coordinates are
\[
\begin{aligned}
	d&=e^{-s}-e^{-4s},\\
	\bar\mu&=\frac{4+2\beta}{3}(1-e^{-3s}),\\
	\omega&=(2+\beta)(e^{-s}-e^{-3s}).
\end{aligned}
\]
All coordinates are nonnegative and below $B$ up to $s_\eps$, and $y\ge\eps$ on this interval.
Thus the path stays strictly inside $D_\eps$.

\paragraph{(C4) The end phase:}
Let $t_1$ be a step at which $Z_{t_1}/n$ is $\eps$-close to $\bar z(s_\eps)$, so that $Y\le 2\eps n$.
Condition on the history up to step $t_1$.
The remaining run can be generated by a uniformly random decision order $\pi$ on $U_{t_1}$, with the same fixed preferences.
First consider the run obtained by removing the remaining pool vertices, while keeping the relative order of all other vertices.
\begin{itemize}
	\item Each triangle and each residual $ab$ edge contributes exactly one edge.
	\item An intact paw contributes two edges if $d$ comes first among its four vertices in $\pi$, and one edge otherwise.
	\item The vertices counted in $D$ or $W$, and any remaining pool pendants, contribute no edges.
\end{itemize}
Let $J$ count the intact paws whose vertex $d$ comes first in $\pi$.
The matching size in this comparison run is $N_T(t_1)+N_2(t_1)+N_I(t_1)+J$.
Restoring the pool adds at most $Y(t_1)$ pool-incident matching edges.
It can also divert $d$ from $c$ in at most $Y(t_1)$ intact paws, changing the internal matching size by at most one in each.
Thus for every $\pi$,
\[
\big|\,|M|-|M_{t_1}|-N_T(t_1)-N_2(t_1)-N_I(t_1)-J\big|\le 2Y(t_1)\le4\eps n.
\]
Conditionally on the history, the relative orders within disjoint paws are independent and uniformly random, so $J\sim\mathrm{Bin}(N_I(t_1),1/4)$.
Since $N_I(t_1)\le n$, Chebyshev's inequality gives $J=N_I(t_1)/4+o(n)$ with probability $1-o(1)$, uniformly in the history.
And therefore (C4) holds with
\[
\Phi=n_T+n_2+\frac54n_I.
\]
The coordinate $n_2$ is zero along the fluid trajectory, but it accounts for any residual $ab$ edges at the cutoff in the discretized run.

\paragraph{The value:}
The edge rate is $g_\mu=(4+2\beta)e^{-3s}$.
At $s^\ast$ the terminal contribution is $\Phi=w^3+\frac14w^4$, so Lemma~\ref{lem:template} gives
\[
\Lambda=\frac{4+2\beta}{3}(1-w^3)+w^3+\frac14w^4=1+\frac14w^4+\frac{1-w^3}{3w^2},
\]
where we used $1+2\beta=w^{-2}$.
Finally, $\nu/n\to 2+\beta=(3w^2+1)/(2w^2)$, giving $F(w)$.
\end{proof}

\begin{proof}[Proof of \cref{thm:rdo-general}]
For $0<w<1$, differentiating \cref{eq:PP-RDO-F(w)} gives us
\[
F'(w)=\frac{w^2p(w)}{(3w^2+1)^2},
\]
where $p(w):=6w^5+3w^3-2w^2-2$.
Ignoring zero coefficients, the coefficient sequence of $p$ has exactly one sign change.
So by Descartes' rule, $p$ has at most one positive root.
Since $p(0)=-2<0<5=p(1)$, there is exactly one such root $w^*\in(0,1)$.
The derivative changes from negative to positive there, so $w^*$ minimizes $F$.
Set $\beta^*=\frac12((w^*)^{-2}-1)$.
Numerically, $w^*\approx0.784470$, $\beta^*\approx0.312489$ and $F(w^*)\approx0.594531$.
Exact rational evaluation gives $p(0.78446)<0<p(0.78447)$, so $0.78446<w^*<0.78447$, and since $w^*$ minimizes $F$, $F(w^*)\le F(0.78447)<0.594532<0.5946$.
Taking $\beta=\beta^*$ in \cref{prop:pawpool}, with $\lfloor\beta^*n\rfloor$ pool edges, the ratio converges to $F(w^*)$.
\end{proof}

\begin{remark}
The construction only uses one common preference order, as in~\cite{TWZ23}; simulations checking the formula are in the repository~\cite{Mukherjee_Fluid_limits_2026}.
\end{remark}

\section{Modified Randomized Greedy}
\label{sec:mrg}

We now return to \MRG{} and to the Double-Bomb graphs recalled in \cref{def:sat}.
Under \MRG, the preferences are uniform, and the satellites interact with the hubs in both directions.
We analyze the two families together, using $\kappa=0$ for $S_n(\sigma)$ and $\kappa=1$ for $S^K_n(\sigma)$.
Throughout this section, $\sigma>0$ is fixed and the number of satellite pairs is $m=\lfloor\sigma n\rfloor$.
We discuss the endpoint $\sigma=0$ after the proof.

\begin{table*}[t]
\caption{Transitions of the \MRG{} chain while both hub classes are nonempty, for actors on the $X$ side.
	We use $\kappa=0$ for $S_n(\sigma)$ and $\kappa=1$ for $S^K_n(\sigma)$.
	Every option creates an edge and increases $|M_t|$ by $1$.
	An actor with no option leaves.
	The $Y$ side is obtained by exchanging $X$ with $Y$, $P$ with $Q$, and $N_{10}$ with $N_{01}$.}
\label{tab:chain}
\begin{tabular}{llll}
	\toprule
	Actor & Number & Option (number) & Change \\
	\midrule
	Hub $x_i$ & $H_X$ & own antenna ($1$) & $H_X,W_X\mathrel{-}=1$ \\
	& & hub of $Y$ ($H_Y$) & $H_X,H_Y\mathrel{-}=1$ \\
	& & other hub of $X$ ($\kappa(H_X-1)$) & $H_X\mathrel{-}=2$ \\
	& & $p_j$ of an intact pair ($N_{11}$) & $H_X,N_{11}\mathrel{-}=1$, $N_{01}\mathrel{+}=1$ \\
	& & $p_j$ with $q_j$ matched ($N_{10}$) & $H_X,N_{10}\mathrel{-}=1$ \\
	Antenna $x'_i$, $x_i$ active & $H_X$ & $x_i$ ($1$) & $H_X,W_X\mathrel{-}=1$ \\
	Antenna $x'_i$, $x_i$ matched & $W_X-H_X$ & none & $W_X\mathrel{-}=1$ \\
	Satellite $p_j$, pair intact & $N_{11}$ & hub of $X$ ($H_X$) & $H_X,N_{11}\mathrel{-}=1$, $N_{01}\mathrel{+}=1$ \\
	& & $q_j$ ($1$) & $N_{11}\mathrel{-}=1$ \\
	Satellite $p_j$, $q_j$ matched & $N_{10}$ & hub of $X$ ($H_X$) & $H_X,N_{10}\mathrel{-}=1$ \\
	\bottomrule
\end{tabular}
\end{table*}

\subsection{Intuition}

In both $S_n(\sigma)$ and $S^K_n(\sigma)$, every hub is eventually matched, so the outcome is decided by who gets the hubs.
Write $h$ for the candidate density of active hubs on each side, and $q$ for the density of active satellites on each side.
Write $\eta=(1+\kappa)h+q$ for the candidate limit of the number of active neighbors of a hub divided by $n$.
Here and below, these variables describe the symmetric fluid trajectory, rather than equalities between the random counts on the two sides.

\paragraph{The race:}
As in \glanceref{}, there are four contributions to the removal rate of an active hub.
It can act itself (rate $1$), its antenna can act (rate $1$), another hub can take it (limiting rate $(1+\kappa)h/\eta$), or a satellite can take it (limiting rate $q/h$).
Here the limiting rates are obtained as $n\to\infty$ along states whose normalized hub and satellite counts approach $h$ and $q$ on each side, with $h>0$.
The total removal rate per hub is at least $3$, whereas the rate per satellite is $1+h/\eta\le 2$.
Thus the satellite-to-hub ratio $r=q/h$ grows.
For $S_n(\sigma)$ it satisfies $r'=r(1+r)$ and diverges at $s^\ast=\log((1+\sigma)/\sigma)$.
The hub density reaches zero at this time, while a positive satellite density remains.
The satellites therefore reduce the opportunity for antennas to match their hubs.

\paragraph{Satellites rarely use their own edge:}
When both hub counts are at least $\eps n$, an acting satellite chooses its own partner with probability at most $1/(\eps n+1)$.
Such choices contribute only $o(n)$ matching edges before this cutoff, with high probability for each fixed $\eps>0$.
The main contribution of satellite--satellite edges comes from the intact pairs left near hub exhaustion, as the end-phase argument below shows.
If $p_j$ matches a hub $x_i$ and $q_j$ remains unmatched at the end, then $x'_i$ is also unmatched.
On these four vertices, the perfect matching $M^*$ contains $x'_ix_i$ and $p_jq_j$, whereas $M$ contains only $x_ip_j$.
The endpoints $x'_i$ and $q_j$ are unmatched, so replacing $x_ip_j$ by the two outer edges of the path $x'_i\,x_i\,p_j\,q_j$ would increase the matching size by one.

\begin{figure}[t]
\centering
\begin{tikzpicture}
	\definecolor{curveS}{HTML}{1F5FA8}\definecolor{curveK}{HTML}{B5530A}\definecolor{curveR}{HTML}{4A8F3C}
	\begin{axis}[
		width=0.98\columnwidth, height=0.72\columnwidth,
		xmin=0, xmax=2.5, ymin=0.628, ymax=0.735,
		axis lines=left, axis line style={-, line width=0.6pt},
		tick style={black, line width=0.5pt}, tick align=outside,
		xtick={0,0.5,1,1.5,2,2.5}, ytick={0.64,0.66,0.68,0.70,0.72},
		yticklabel style={/pgf/number format/fixed, /pgf/number format/precision=2, /pgf/number format/fixed zerofill},
		xticklabel style={/pgf/number format/fixed, /pgf/number format/precision=1, /pgf/number format/fixed zerofill},
		xlabel={satellite density $\sigma$}, ylabel={limiting ratio $\E|M|/\nu$},
		label style={font=\normalsize}, tick label style={font=\normalsize},
		legend style={draw=none, fill=none, at={(1,1)}, anchor=north east, font=\normalsize,
			row sep=-1pt, cells={anchor=west}},
		legend cell align=left,
		every axis plot/.append style={line width=1.4pt},
		clip=false]
		\addplot[gray!70, dashed, line width=0.8pt, forget plot] coordinates {(0,0.666667) (2.5,0.666667)};
		\node[gray!90!black, anchor=north east, font=\normalsize] at (axis cs:2.48,0.6655) {bomb, $2/3$};
		\addplot[curveS] coordinates {
			(0.0000,0.66667) (0.0201,0.66990) (0.0402,0.67288) (0.0602,0.67558) (0.0803,0.67797) (0.1004,0.68007)
			(0.1205,0.68187) (0.1406,0.68338) (0.1606,0.68462) (0.1807,0.68561) (0.2008,0.68635) (0.2209,0.68688)
			(0.2410,0.68720) (0.2610,0.68734) (0.2811,0.68732) (0.3012,0.68716) (0.3213,0.68686) (0.3414,0.68645)
			(0.3614,0.68595) (0.3815,0.68535) (0.4016,0.68469) (0.4217,0.68396) (0.4418,0.68318) (0.4618,0.68236)
			(0.4819,0.68151) (0.5020,0.68064) (0.5221,0.67974) (0.5422,0.67884) (0.5622,0.67794) (0.5823,0.67703)
			(0.6024,0.67613) (0.6225,0.67524) (0.6426,0.67436) (0.6627,0.67349) (0.6827,0.67265) (0.7028,0.67183)
			(0.7229,0.67103) (0.7430,0.67026) (0.7631,0.66951) (0.7831,0.66879) (0.8032,0.66810) (0.8233,0.66744)
			(0.8434,0.66681) (0.8635,0.66621) (0.8835,0.66564) (0.9036,0.66511) (0.9237,0.66460) (0.9438,0.66413)
			(0.9639,0.66369) (0.9839,0.66329) (1.0040,0.66291) (1.0241,0.66256) (1.0442,0.66225) (1.0643,0.66196)
			(1.0843,0.66171) (1.1044,0.66148) (1.1245,0.66129) (1.1446,0.66112) (1.1647,0.66098) (1.1847,0.66087)
			(1.2048,0.66078) (1.2249,0.66072) (1.2450,0.66069) (1.2651,0.66068) (1.2851,0.66069) (1.3052,0.66073)
			(1.3253,0.66079) (1.3454,0.66088) (1.3655,0.66098) (1.3855,0.66111) (1.4056,0.66126) (1.4257,0.66143)
			(1.4458,0.66162) (1.4659,0.66182) (1.4859,0.66205) (1.5060,0.66229) (1.5261,0.66255) (1.5462,0.66283)
			(1.5663,0.66313) (1.5863,0.66344) (1.6064,0.66376) (1.6265,0.66410) (1.6466,0.66446) (1.6667,0.66483)
			(1.6867,0.66521) (1.7068,0.66560) (1.7269,0.66601) (1.7470,0.66643) (1.7671,0.66686) (1.7871,0.66730)
			(1.8072,0.66776) (1.8273,0.66822) (1.8474,0.66869) (1.8675,0.66918) (1.8876,0.66967) (1.9076,0.67017)
			(1.9277,0.67068) (1.9478,0.67120) (1.9679,0.67173) (1.9880,0.67226) (2.0080,0.67280) (2.0281,0.67335)
			(2.0482,0.67391) (2.0683,0.67447) (2.0884,0.67504) (2.1084,0.67561) (2.1285,0.67619) (2.1486,0.67678)
			(2.1687,0.67737) (2.1888,0.67797) (2.2088,0.67857) (2.2289,0.67917) (2.2490,0.67978) (2.2691,0.68040)
			(2.2892,0.68102) (2.3092,0.68164) (2.3293,0.68226) (2.3494,0.68289) (2.3695,0.68352) (2.3896,0.68416)
			(2.4096,0.68480) (2.4297,0.68544) (2.4498,0.68608) (2.4699,0.68673) (2.4900,0.68737) (2.5000,0.68770)
		};
		\addlegendentry{\MRG{} on $S_n(\sigma)$ (bipartite)}
		\addplot[curveK] coordinates {
			(0.0000,0.66667) (0.0201,0.66867) (0.0402,0.67029) (0.0602,0.67157) (0.0803,0.67258) (0.1004,0.67333)
			(0.1205,0.67387) (0.1406,0.67422) (0.1606,0.67440) (0.1807,0.67444) (0.2008,0.67435) (0.2209,0.67416)
			(0.2410,0.67387) (0.2610,0.67350) (0.2811,0.67307) (0.3012,0.67258) (0.3213,0.67205) (0.3414,0.67148)
			(0.3614,0.67087) (0.3815,0.67025) (0.4016,0.66961) (0.4217,0.66896) (0.4418,0.66831) (0.4618,0.66765)
			(0.4819,0.66700) (0.5020,0.66635) (0.5221,0.66571) (0.5422,0.66509) (0.5622,0.66448) (0.5823,0.66388)
			(0.6024,0.66331) (0.6225,0.66275) (0.6426,0.66222) (0.6627,0.66170) (0.6827,0.66121) (0.7028,0.66075)
			(0.7229,0.66031) (0.7430,0.65989) (0.7631,0.65950) (0.7831,0.65913) (0.8032,0.65879) (0.8233,0.65848)
			(0.8434,0.65819) (0.8635,0.65793) (0.8835,0.65770) (0.9036,0.65748) (0.9237,0.65730) (0.9438,0.65714)
			(0.9639,0.65701) (0.9839,0.65689) (1.0040,0.65681) (1.0241,0.65675) (1.0442,0.65671) (1.0643,0.65669)
			(1.0843,0.65670) (1.1044,0.65673) (1.1245,0.65678) (1.1446,0.65685) (1.1647,0.65694) (1.1847,0.65705)
			(1.2048,0.65718) (1.2249,0.65733) (1.2450,0.65750) (1.2651,0.65769) (1.2851,0.65790) (1.3052,0.65812)
			(1.3253,0.65836) (1.3454,0.65862) (1.3655,0.65889) (1.3855,0.65918) (1.4056,0.65949) (1.4257,0.65981)
			(1.4458,0.66014) (1.4659,0.66049) (1.4859,0.66085) (1.5060,0.66122) (1.5261,0.66161) (1.5462,0.66201)
			(1.5663,0.66242) (1.5863,0.66284) (1.6064,0.66327) (1.6265,0.66371) (1.6466,0.66417) (1.6667,0.66463)
			(1.6867,0.66511) (1.7068,0.66559) (1.7269,0.66608) (1.7470,0.66658) (1.7671,0.66709) (1.7871,0.66761)
			(1.8072,0.66814) (1.8273,0.66867) (1.8474,0.66921) (1.8675,0.66976) (1.8876,0.67031) (1.9076,0.67087)
			(1.9277,0.67144) (1.9478,0.67201) (1.9679,0.67259) (1.9880,0.67317) (2.0080,0.67376) (2.0281,0.67436)
			(2.0482,0.67495) (2.0683,0.67556) (2.0884,0.67617) (2.1084,0.67678) (2.1285,0.67739) (2.1486,0.67801)
			(2.1687,0.67864) (2.1888,0.67927) (2.2088,0.67990) (2.2289,0.68053) (2.2490,0.68117) (2.2691,0.68181)
			(2.2892,0.68245) (2.3092,0.68309) (2.3293,0.68374) (2.3494,0.68439) (2.3695,0.68504) (2.3896,0.68569)
			(2.4096,0.68635) (2.4297,0.68700) (2.4498,0.68766) (2.4699,0.68832) (2.4900,0.68898) (2.5000,0.68931)
		};
		\addlegendentry{\MRG{} on $S^K_n(\sigma)$}
		\addplot[curveR] coordinates {
			(0.0000,0.66667) (0.0201,0.66610) (0.0402,0.66521) (0.0602,0.66419) (0.0803,0.66310) (0.1004,0.66198)
			(0.1205,0.66087) (0.1406,0.65976) (0.1606,0.65868) (0.1807,0.65763) (0.2008,0.65662) (0.2209,0.65565)
			(0.2410,0.65472) (0.2610,0.65384) (0.2811,0.65301) (0.3012,0.65223) (0.3213,0.65150) (0.3414,0.65081)
			(0.3614,0.65017) (0.3815,0.64958) (0.4016,0.64904) (0.4217,0.64855) (0.4418,0.64810) (0.4618,0.64769)
			(0.4819,0.64733) (0.5020,0.64701) (0.5221,0.64673) (0.5422,0.64649) (0.5622,0.64629) (0.5823,0.64613)
			(0.6024,0.64600) (0.6225,0.64591) (0.6426,0.64586) (0.6627,0.64583) (0.6827,0.64584) (0.7028,0.64588)
			(0.7229,0.64595) (0.7430,0.64605) (0.7631,0.64618) (0.7831,0.64634) (0.8032,0.64652) (0.8233,0.64672)
			(0.8434,0.64695) (0.8635,0.64720) (0.8835,0.64748) (0.9036,0.64778) (0.9237,0.64809) (0.9438,0.64843)
			(0.9639,0.64879) (0.9839,0.64916) (1.0040,0.64956) (1.0241,0.64997) (1.0442,0.65040) (1.0643,0.65084)
			(1.0843,0.65130) (1.1044,0.65177) (1.1245,0.65225) (1.1446,0.65275) (1.1647,0.65327) (1.1847,0.65379)
			(1.2048,0.65433) (1.2249,0.65488) (1.2450,0.65544) (1.2651,0.65600) (1.2851,0.65658) (1.3052,0.65717)
			(1.3253,0.65777) (1.3454,0.65838) (1.3655,0.65899) (1.3855,0.65961) (1.4056,0.66024) (1.4257,0.66088)
			(1.4458,0.66153) (1.4659,0.66218) (1.4859,0.66283) (1.5060,0.66350) (1.5261,0.66416) (1.5462,0.66484)
			(1.5663,0.66552) (1.5863,0.66620) (1.6064,0.66689) (1.6265,0.66758) (1.6466,0.66828) (1.6667,0.66898)
			(1.6867,0.66968) (1.7068,0.67039) (1.7269,0.67110) (1.7470,0.67181) (1.7671,0.67252) (1.7871,0.67324)
			(1.8072,0.67396) (1.8273,0.67468) (1.8474,0.67541) (1.8675,0.67614) (1.8876,0.67686) (1.9076,0.67759)
			(1.9277,0.67832) (1.9478,0.67906) (1.9679,0.67979) (1.9880,0.68052) (2.0080,0.68126) (2.0281,0.68199)
			(2.0482,0.68273) (2.0683,0.68347) (2.0884,0.68420) (2.1084,0.68494) (2.1285,0.68568) (2.1486,0.68642)
			(2.1687,0.68716) (2.1888,0.68789) (2.2088,0.68863) (2.2289,0.68937) (2.2490,0.69010) (2.2691,0.69084)
			(2.2892,0.69158) (2.3092,0.69231) (2.3293,0.69305) (2.3494,0.69378) (2.3695,0.69451) (2.3896,0.69524)
			(2.4096,0.69597) (2.4297,0.69670) (2.4498,0.69743) (2.4699,0.69816) (2.4900,0.69889) (2.5000,0.69925)
		};
		\addlegendentry{\RDO{} (hub-first) on $S_n(\sigma)$}
		\addplot[curveS, only marks, mark=*, mark size=1.6pt, forget plot] coordinates {(1.2626108,0.660675593)};
		\addplot[curveK, only marks, mark=*, mark size=1.6pt, forget plot] coordinates {(1.0685868,0.656689832)};
		\addplot[curveR, only marks, mark=*, mark size=1.6pt, forget plot] coordinates {(0.666667,0.645833)};
		\draw[curveR, line width=0.6pt] (axis cs:0.36,0.6352) -- (axis cs:0.6667,0.64583);
		\node[curveR, anchor=west, font=\normalsize, inner sep=1pt] at (axis cs:0.06,0.6345) {$31/48$};
	\end{axis}
\end{tikzpicture}
\Description{Three curves of the limiting matching ratio against the satellite density sigma. The two MRG curves initially rise above two thirds and then fall below it near sigma equal to one. The RDO curve has its minimum at sigma equal to two thirds.}
\caption{Limiting expected matching ratios on the Double-Bomb families.
	The \MRG{} curves are numerical evaluations of the full integral formulas in \cref{eq:mrg-bip-integral,eq:mrg-clique-integral}, and the \RDO{} curve is evaluated from the exact expression in \cref{thm:rdo-bip}.
	Dots mark the numerically located minima $0.66068$ at $\sigma\approx1.26$ for \MRG{} on $S_n(\sigma)$ and $0.65669$ at $\sigma\approx1.07$ for \MRG{} on $S^K_n(\sigma)$, and the exact minimum $31/48$ at $\sigma=2/3$ for \RDO{} on $S_n(\sigma)$.
	All three ratios equal $2/3$ at $\sigma=0$, and the two \MRG{} curves initially increase.}
\label{fig:ratio}
\end{figure}
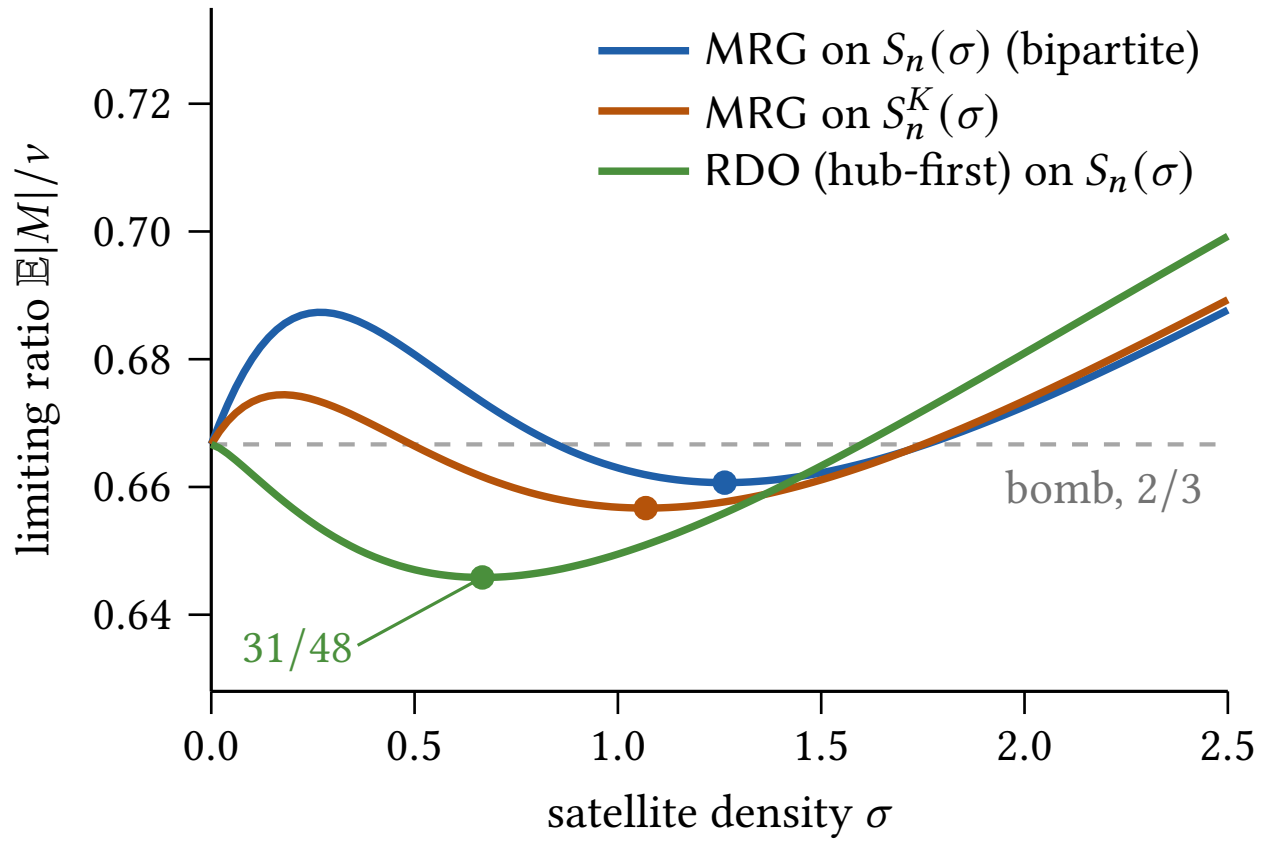

\paragraph{A trade-off:}
Adding satellites has two competing effects.
They give hubs an alternative to hub--hub matches, but they also take hubs away from the antennas.
At densities around $1$, the ratio falls below $2/3$, as shown in \cref{fig:ratio}.

\subsection{The Fluid Limit}

The rates above give the following system in log-time $s$, with initial conditions $h(0)=1$ and $q(0)=\sigma$.
\begin{equation}
\label{eq:satode}
\begin{aligned}
	h'&=-h\Big(2+\frac{(1+\kappa)h}{\eta}\Big)-q,\\
	q'&=-q\Big(1+\frac{h}{\eta}\Big),
	\text{ where }\eta=(1+\kappa)h+q.
\end{aligned}
\end{equation}
While both hub classes remain nonempty, every action of a hub, an antenna attached to a hub, or a satellite creates an edge.
The normalized edge rate is therefore $4h+2q$.

\begin{lemma}[Fluid limit of the Double-Bomb graphs]
\label{lem:satfluid}
Fix $\sigma>0$ and $\kappa\in\{0,1\}$, and let $G_n$ be the corresponding Double-Bomb graph.
Let $(h,q)$ solve \cref{eq:satode}, and let $s^\ast$ be the first time at which $h=0$.
Then $s^\ast<\infty$, $q(s^\ast)>0$, and $|M(G_n)|/n\to\Lambda_\kappa(\sigma)$ in probability, where
\[
\Lambda_\kappa(\sigma)=\int_0^{s^\ast}(4h+2q)\,ds+\frac{q(s^\ast)^2}{\sigma}.
\]
Consequently, $\E|M(G_n)|/\nu(G_n)\to\Lambda_\kappa(\sigma)/(2+\sigma)$.
\end{lemma}

\begin{proof}
We check the conditions of \cref{lem:template}, treating both families at once.

\paragraph{(C1) The chain:}
We use the same count vector $Z$ as in the proof of \cref{thm:rdo-bip},
\[
Z=(H_X,H_Y,W_X,W_Y,N_{11},N_{10},N_{01},|M_t|).
\]
Every active hub still has an active antenna.
In \cref{tab:chain}, the actor class is chosen with probability proportional to its active count, and the partner class is then chosen with probability proportional to its available count.
The transitions depend only on $Z$, and each coordinate changes by at most $2$.

Equivalently, we can permute the indices $[n]$ simultaneously on $X$ and $X'$, independently do the same on $Y$ and $Y'$, and permute $[m]$ simultaneously on $P$ and $Q$.
These are graph automorphisms, and they map any reachable active set to any other with the same counts.
Since \MRG{} is invariant under these automorphisms, the counts determine the conditional law of the next step.

\paragraph{(C2) The drift:}
For hubs in $X$ and $Y$, the numbers of active neighbors divided by $n$ are, up to an $O(1/n)$ correction,
\[
\begin{aligned}
	\eta_X&=h_Y+n_{11}+n_{10}+\kappa h_X,\\
	\eta_Y&=h_X+n_{11}+n_{01}+\kappa h_Y.
\end{aligned}
\]
The exact number of active neighbors of an $X$-hub is $n\eta_X+1-\kappa$, and similarly for a $Y$-hub.
To define $g$, we keep the leading terms of the drift as $n\to\infty$.
When both hub counts exceed $\eps n/2$, an acting hub chooses its own antenna, or an acting satellite chooses its own partner, with probability $O(1/(\eps n))$.
These choices remain in the discrete process, but their contributions vanish from the limiting rates.
We therefore omit them from $g$ and replace $\kappa(H_X-1)$ by $\kappa H_X$, and similarly on the $Y$ side.
The comparison below bounds the resulting error in the conditional expected increment of $Z_t$ by $C_\eps/n$, as required by (C2).
The resulting log-time rates are
\begin{equation}
	\label{eq:mrg-drift}
	\begin{aligned}
		g_{h_X}&=-2h_X-\frac{\kappa h_X^2}{\eta_X}-\frac{h_Xh_Y}{\eta_Y}-n_{11}-n_{10},\\
		g_{w_X}&=-w_X,\\
		g_{n_{11}}&=-n_{11}\Big(2+\frac{h_X}{\eta_X}+\frac{h_Y}{\eta_Y}\Big),\\
		g_{n_{10}}&=n_{11}\Big(1+\frac{h_Y}{\eta_Y}\Big)-n_{10}\Big(1+\frac{h_X}{\eta_X}\Big),\\
		g_\mu&=2h_X+2h_Y+2n_{11}+n_{10}+n_{01}.
	\end{aligned}
\end{equation}
The remaining coordinates follow by exchanging $X$ with $Y$ and $n_{10}$ with $n_{01}$.

Let $B=5+2\sigma$, and let $D_\eps$ be the box where $h_X,h_Y\in(\eps/2,B)$ and $w_X,w_Y,n_{11},n_{10},n_{01},\mu\in(-\eps/8,B)$.
On this box, $\eta_X,\eta_Y>\eps/4$, and the active-vertex mass satisfies $u>\eps/4$.

Each discrete transition probability is the product of an actor probability and a conditional partner probability.
For example, an acting hub of $X$ chooses a hub of $Y$ with probability
\[
\frac{H_Y}{1+H_Y+\kappa(H_X-1)+N_{11}+N_{10}}.
\]
The approximation uses $h_Y/\eta_X$, whose denominator before rescaling is $H_Y+\kappa H_X+N_{11}+N_{10}$.
The two denominators differ by $1-\kappa$, and both are at least $\eps n/4$ in the states under consideration.
The same-side hub numerator changes by at most $1$, and the omitted antenna and partner options each have probability $O(1/(\eps n))$.
Thus each of the $20$ transition probabilities, including the omitted transitions with approximate probability zero, changes by at most $8/(\eps n)$.
Since every coordinate increment has absolute value at most $2$, (C2) holds with $C_\eps=400/\eps$.
The map $g$ is rational on the bounded box, with all denominators bounded away from zero, so it is Lipschitz.

\paragraph{(C3) The solution:}
The ODE with rates given in \cref{eq:mrg-drift} and initial condition $z_0=(1,1,1,1,\sigma,0,0,0)$ is symmetric under exchanging the two sides.
By uniqueness, $h_X=h_Y=h$, $n_{10}=n_{01}$ and $\eta_X=\eta_Y=\eta$ along the deterministic solution.
Setting $q=n_{11}+n_{10}$ gives \cref{eq:satode}.
Also $w_X=w_Y=e^{-s}$ and $n_{11}=q^2/\sigma$, since both $n_{11}$ and $q^2/\sigma$ solve $b'=-2(1+h/\eta)b$ with initial value $\sigma$.

For $r=q/h$, \cref{eq:satode} gives
\[
r'=r\Big(1+r+\frac{\kappa}{1+\kappa+r}\Big)\ge r(1+r).
\]
Thus $r$ diverges at a finite time $s^\ast\le\log((1+\sigma)/\sigma)$.
While $h>0$, we have $\sigma e^{-2s}\le q(s)\le\sigma e^{-s}$.
It follows that $h=q/r$ tends to zero at $s^\ast$, and $q(s^\ast)>0$.

Take the budget to be $h_X$, and let $s_\eps$ be the first time at which $h=\eps$.
It exists and tends to $s^\ast$, since $h'<0$ while $h>0$.
Up to this time, the hub densities lie in $[\eps,1]$, the antenna densities lie in $[0,1]$, and the satellite-pair densities lie in $[0,\sigma]$.
In particular, $n_{10}=n_{01}=q(1-q/\sigma)\ge0$.
Finally, $\bar\mu'=4h+2q\le-2h'$, so $\bar\mu\le2(1-h)\le2$.
Hence the trajectory stays strictly inside $D_\eps$.

\paragraph{(C4) The end phase:}
The argument from the proof of \cref{thm:rdo-bip} applies to both families.
Every edge is either incident to a hub or is a satellite-pair edge, including when the hub classes are cliques.
If the state at $t_1$ is $\eps$-close to $\bar z(s_\eps)$, at most $4\eps n$ active hubs remain.
Maximality of the output matching therefore gives
\[
\big|\,|M|-|M_{t_1}|-N_{11}(t_1)\big|\le4\eps n.
\]
So, (C4) holds with $\Phi=n_{11}$.
Since $n_{11}(s)=q(s)^2/\sigma$, Lemma~\ref{lem:template} gives the stated limit.
\end{proof}

When $\sigma=0$, there are no satellites.
For either value of $\kappa$, the symmetric hub equation reduces to $h'=-3h$ and the edge rate is $4h$.
The same cutoff argument applies with $s_\eps=\frac13\log(1/\eps)$ and $\Phi=0$.
It gives $\Lambda_\kappa(0)=4/3$ and limiting ratio $2/3$ for both families.
For $\kappa=0$, this recovers the Bomb proof in \cref{sec:bomb-proof}.

\subsection{Solving the Fluid Equations}
\label{sec:mrg-closed}

The equation for $r=q/h$ lets us solve the two cases explicitly.

\paragraph{The bipartite family ($\kappa=0$):}
For $\kappa=0$, we have $r'=r(1+r)$ with $r(0)=\sigma$.
Writing $v=e^s$ gives $r=\sigma v/(1+\sigma-\sigma v)$ and
\[
\begin{aligned}
h(v)&=v^{-3}(1+\sigma-\sigma v)e^{\sigma(v-1)/(1+\sigma)},\\
q(v)&=\sigma v^{-2}e^{\sigma(v-1)/(1+\sigma)}.
\end{aligned}
\]
Since $v=e^s$, the time interval $0\le s\le s^\ast$ becomes $1\le v\le(1+\sigma)/\sigma$.
Thus the upper integration limit below is $e^{s^\ast}=(1+\sigma)/\sigma$, the value of $v$ at hub exhaustion.
Substituting in \cref{lem:satfluid} yields
\begin{equation}
\label{eq:mrg-bip-integral}
\begin{aligned}
	\Lambda_0(\sigma)
	&=\int_1^{(1+\sigma)/\sigma}\Big(\frac{4(1+\sigma)}{v^4}-\frac{2\sigma}{v^3}\Big)
	e^{\sigma(v-1)/(1+\sigma)}\, d v\\
	& +\sigma\Big(\frac{\sigma}{1+\sigma}\Big)^4e^{2/(1+\sigma)}.
\end{aligned}
\end{equation}

\paragraph{The general family ($\kappa=1$):}
For the case $\kappa=1$, define
\[
\gamma(r):=r^2+3r+3 \text{ and } \theta(r):=\arctan\frac{2r+3}{\sqrt3}.
\]
Then $r'=r\gamma(r)/(r+2)$, and
\[
\frac{d\log h}{dr}=-\frac{r^2+4r+6}{r\gamma(r)}.
\]
Integrating with the initial condition $h(\sigma)=1$ gives
\[
h(r)=\Big(\frac\sigma r\Big)^2
\sqrt{\frac{\gamma(r)}{\gamma(\sigma)}}\,
e^{(\theta(r)-\theta(\sigma))/\sqrt3}.
\]
In particular, the terminal satellite density is
\[
q_\ast:=q(s^\ast)=
\frac{\sigma^2}{\sqrt{\gamma(\sigma)}}
e^{(\pi/2-\theta(\sigma))/\sqrt3}.
\]
Using $ds=(r+2)\,dr/(r\gamma(r))$, we obtain
\begin{equation}
\label{eq:mrg-clique-integral}
\Lambda_1(\sigma)=\int_\sigma^\infty
h(r)\frac{2(r+2)^2}{r\gamma(r)}\, d r+\frac{q_\ast^2}{\sigma}.
\end{equation}

\subsection{Proofs of \texorpdfstring{\cref{thm:mrg-bip,thm:mrg-general}}{the MRG bounds}}

By \cref{lem:satfluid}, the limits as $n\to\infty$ of the expected matching ratios $\E|M|/\nu$ on the two families are
\[
\rho_S(\sigma)=\frac{\Lambda_0(\sigma)}{2+\sigma}
\text{ and }
\rho_K(\sigma)=\frac{\Lambda_1(\sigma)}{2+\sigma}.
\]
The exact constants in \cref{thm:mrg-bip,thm:mrg-general} are therefore given by \cref{eq:mrg-bip-integral,eq:mrg-clique-integral} at $\sigma=5/4$ and $\sigma=1$, respectively.
Their numerical values are
\begin{align*}
\rho_S(5/4)&=0.6606804\ldots,\\
\rho_K(1)&=0.6568236\ldots.
\end{align*}
For the stated decimal upper bounds, monotone Riemann sums give
$\rho_S(5/4)<0.660698381<0.6607$ and
$\rho_K(1)<0.656892555<0.6569$.
\footnote{For \cref{eq:mrg-bip-integral} at $\sigma=5/4$, the integrand is decreasing on $[1,9/5]$, so a left sum with $40{,}000$ equal subintervals gives an upper bound.
For \cref{eq:mrg-clique-integral} at $\sigma=1$, substituting $t=1/r$ gives an integral on $[0,1]$ with positive integrand $J$ satisfying
$J'(t)/J(t)=(2+5t+6t^2)/((1+2t)(1+3t+3t^2))>0$.
A right sum with $4{,}000$ equal subintervals gives the second bound.
The elementary functions in these finite sums can be enclosed using Taylor remainder bounds.}
This proves \cref{thm:mrg-bip,thm:mrg-general}.
\qed

\section{Conclusion and the Behavior Near the Bomb Graph}
\label{sec:conclusion}
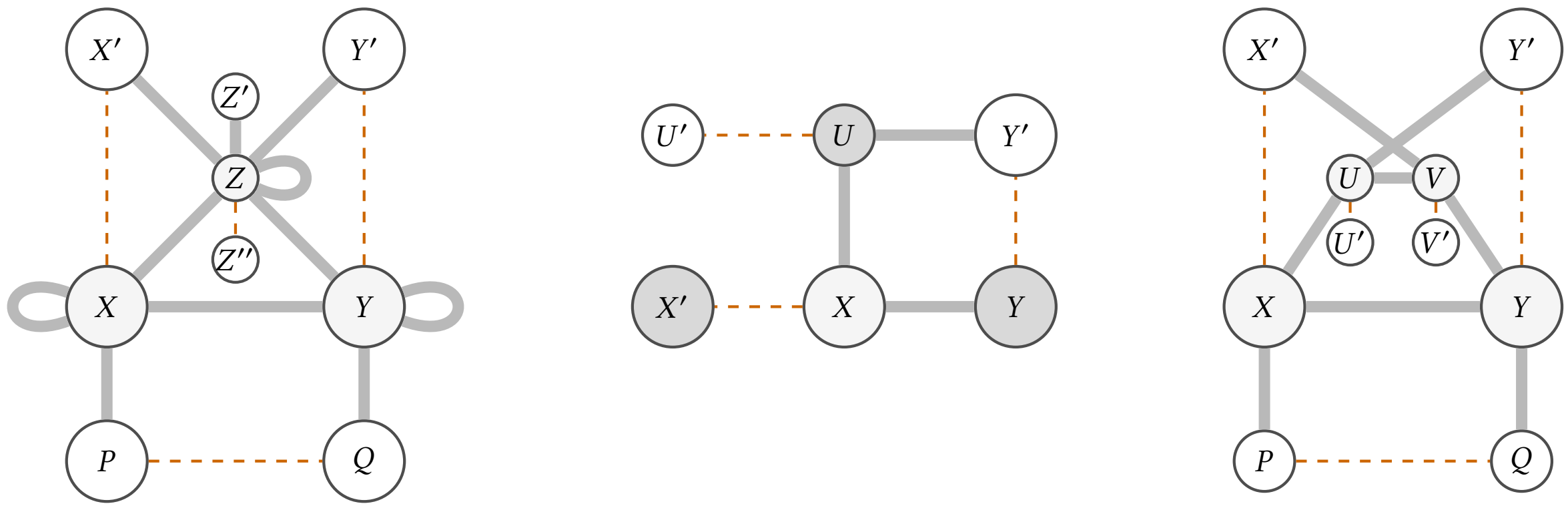
\begin{figure*}[t]
\centering
\begin{tikzpicture}[x=0.85cm, y=0.85cm,
	cls/.style={circle, draw=black!70, fill=gray!8, thick, minimum size=0.8cm, inner sep=0pt},
	ant/.style={circle, draw=black!70, fill=white, thick, minimum size=0.8cm, inner sep=0pt},
	sat/.style={circle, draw=black!70, fill=white, thick, minimum size=0.6cm, inner sep=0pt},
	tiny/.style={circle, draw=black!70, fill=white, thick, minimum size=0.45cm, inner sep=0pt},
	tinycls/.style={tiny, fill=gray!8},
	sidea/.style={fill=gray!30},
	blk/.style={line width=3.2pt, draw=gray!55},
	opt/.style={thick, dashed, draw=orange!80!black},
	sparse/.style={thick, draw=black},
	lab/.style={font=\normalsize, align=center}]
	\begin{scope}
		\node[ant] (Xp) at (0,3) {$X'$};   \node[ant] (Yp) at (3,3) {$Y'$};
		\node[cls] (X) at (0,0) {$X$};     \node[cls] (Y) at (3,0) {$Y$};
		\node[ant] (P) at (0,-1.8) {$P$};  \node[ant] (Q) at (3,-1.8) {$Q$};
		\node[tinycls] (Z) at (1.5,1.5) {$Z$};
		\node[tiny] (Z1) at (1.5,2.45) {$Z'$};
		\node[tiny] (Z2) at (1.5,0.55) {$Z''$};
		\draw[blk] (X) -- (Y); \draw[blk] (X) -- (P); \draw[blk] (Y) -- (Q);
		\draw[blk] (X) to[out=200,in=160,looseness=7] (X);
		\draw[blk] (Y) to[out=20,in=-20,looseness=7] (Y);
		\foreach \u in {Xp,X,Yp,Y,Z1}{\draw[blk] (Z) -- (\u);}
		\draw[blk] (Z) to[out=-25,in=25,looseness=9] (Z);
		\draw[opt] (X) -- (Xp); \draw[opt] (Y) -- (Yp); \draw[opt] (P) -- (Q); \draw[opt] (Z) -- (Z2);
		\node[lab] at (1.5,-3.05) {(a) \MRG{}, general graphs, $\approx0.6555$};
		\node[lab] at (1.5,-3.75) {$|Y|=|Y'|=0.827n$,\ $|Z|=|Z'|=|Z''|=0.062n$};
	\end{scope}
	\begin{scope}[xshift=6.6*0.85cm]
		\node[ant, sidea] (Xp) at (0,0) {$X'$};
		\node[cls] (X) at (2,0) {$X$};
		\node[cls, sidea] (Y) at (4,0) {$Y$};
		\node[sat] (U1) at (0,2) {$U'$};
		\node[sat, sidea] (U) at (2,2) {$U$};
		\node[ant] (Yp) at (4,2) {$Y'$};
		\draw[blk] (X) -- (Y); \draw[blk] (U) -- (X); \draw[blk] (U) -- (Yp);
		\draw[opt] (X) -- (Xp); \draw[opt] (Y) -- (Yp); \draw[opt] (U) -- (U1);
		\node[lab] at (2,-3.05) {(b) \MRG{}, bipartite graphs, $\approx0.6603$};
		\node[lab] at (2,-3.75) {$|U|=|U'|=0.454n$};
	\end{scope}
	\begin{scope}[xshift=13.5*0.85cm]
		\coordinate (cXp) at (0,3); \coordinate (cYp) at (3,3);
		\coordinate (cX) at (0,0);  \coordinate (cY) at (3,0);
		\coordinate (cM) at ($(cX)!0.5!(cXp)$);
		\coordinate (cU) at ($(cM)+1/3*(cY)-1/3*(cX)$);
		\coordinate (cV) at ($(cM)+2/3*(cY)-2/3*(cX)$);
		\coordinate (cU1) at ($(cU)-1/4*(cXp)+1/4*(cX)$);
		\coordinate (cV1) at ($(cV)-1/4*(cXp)+1/4*(cX)$);
		\node[ant] (Xp) at (cXp) {$X'$};   \node[ant] (Yp) at (cYp) {$Y'$};
		\node[cls] (X) at (cX) {$X$};      \node[cls] (Y) at (cY) {$Y$};
		\node[sat] (P) at (0,-1.8) {$P$};  \node[sat] (Q) at (3,-1.8) {$Q$};
		\node[tinycls] (U) at (cU) {$U$};  \node[tiny] (U1) at (cU1) {$U'$};
		\node[tinycls] (V) at (cV) {$V$};  \node[tiny] (V1) at (cV1) {$V'$};
		\draw[blk] (X) -- (Y); \draw[blk] (X) -- (P); \draw[blk] (Y) -- (Q);
		\draw[blk] (U) -- (X); \draw[blk] (U) -- (Yp); \draw[blk] (U) -- (V);
		\draw[blk] (V) -- (Xp); \draw[blk] (V) -- (Y);
		\draw[opt] (X) -- (Xp); \draw[opt] (Y) -- (Yp); \draw[opt] (P) -- (Q);
		\draw[opt] (U) -- (U1); \draw[opt] (V) -- (V1);
		\node[lab] at (1.5,-3.05) {(c) \RDO{}, bipartite graphs, $\approx0.6431$};
		\node[lab] at (1.5,-3.75) {$|Y|=|Y'|=1.006n$,\ $|P|=|Q|=0.705n$,};
		\node[lab] at (1.5,-4.35) {$|U|=|U'|=0.063n$,\ $|V|=|V'|=0.106n$};
	\end{scope}
\end{tikzpicture}
\Description{Three class-level diagrams in the style of Figure 1. (a) The Double-Bomb with cliques and unequal sides, with a small third hub class Z in the middle joined to X', X, Y, Y' and to its pendant class Z'. (b) A ladder: the bomb graph X'--X--Y--Y' with an extra class U joined to X and Y', each vertex of U with a private pendant in U'. (c) The Double-Bomb with two small added pairs U U' and V V'.}
\caption{Numerically promising extensions of the Bomb and Double-Bomb templates, found by the local search of \cref{sec:toolkit-search}.
	Conventions are as in \cref{fig:constructions}, and classes have $n$ vertices unless stated.
	The ratios are numerical estimates for the stated weights and preferences, not bounds proved in this paper.
	In (b) the shaded classes form one side of the bipartition.
	In (c) the \RDO{} preferences are, from most to least preferred, with a uniformly random choice inside braces (``own'' is the in-copy partner):
	$X\colon Y\succ P\succ\text{own}\succ U$;
	$X'\colon\{\text{own},V\}$;
	$Y\colon V\succ X\succ Q\succ\text{own}$;
	$Y'\colon U\succ\text{own}$;
	$P\colon X\succ\text{own}$;
	$Q\colon Y\succ\text{own}$;
	$U\colon X\succ\{Y',\text{own},V\}$;
	$V\colon\{\text{own},Y,U\}\succ X'$;
	$U',V'\colon\text{own}$.}
\label{fig:candidates}
\end{figure*}

In this work, we showed that \MRG{} is not $2/3$-approximate, on general or on bipartite graphs, and we improved the upper bounds for \RDO{} on both classes.
Our proofs follow a common structure.
A finite count vector describes the discrete run, Wormald's ODE method controls its trajectory while the relevant classes have positive density, and a separate end-phase argument determines the remaining contribution.
We conclude by examining why small perturbations of the Double-Bomb families do not improve the classical \MRG{} bound, and by discussing open questions.

\paragraph{$\rho_S(\sigma), \rho_K(\sigma)$ have a local minimum at $\sigma=0$:}
The behavior near $\sigma=0$ gives one possible explanation for why the bound $2/3$ looked optimal.
The closed forms above have the following expansions.

\begin{proposition}[Expansions at $\sigma=0$]
\label{prop:expansions}
As $\sigma\downarrow0$,
\begin{align*}
	\rho_S(\sigma)&=\frac23+\frac\sigma6+O(\sigma^2),\\
	\rho_K(\sigma)&=\frac23+\frac\sigma9-\frac{\sigma^2}{27}\log\frac1\sigma+O(\sigma^2),\\
	\rho_{\RDO}(\sigma)&=\frac23-\frac{\sqrt2}{6}\,\sigma^{3/2}+\frac{\sigma^2}4+O(\sigma^{5/2}).
\end{align*}
\end{proposition}

\begin{proof}[Proof]
The \RDO{} expansion follows by substituting $w^2=\sigma/(2+\sigma)$ into $\frac23-\frac23w^3+w^4$.
For $\rho_S$, put $c=\sigma/(1+\sigma)$ and expand the exponential in \cref{eq:mrg-bip-integral} as $1+c(v-1)+O(c^2(v-1)^2)$, uniformly since $0\le c(v-1)\le1$ on the integration interval.
As $\int_1^\infty v^{-4}(v-1)^2\,dv<\infty$, the remainder is $O(\sigma^2)$, the terminal term is $O(\sigma^5)$, and $\Lambda_0(\sigma)=\frac43+\sigma+O(\sigma^2)$.
For $\rho_K$, extract the prefactor $\sigma^2e^{-\theta(\sigma)/\sqrt3}/\sqrt{\gamma(\sigma)}$ from the integral in \cref{eq:mrg-clique-integral}, split the rest at $r=1$, and expand the integrand through its $r^{-1}$ term at zero and the prefactor to first order.
The remainders on $[\sigma,1]$ and $[1,\infty)$ are $O(\sigma^2)$, the terminal term is $O(\sigma^3)$, and the $r^{-1}$ term gives $\Lambda_1(\sigma)=\frac43+\frac89\sigma-\frac2{27}\sigma^2\log\frac1\sigma+O(\sigma^2)$.
Dividing by $2+\sigma$ gives the two \MRG{} expansions; the script \texttt{verify\_sigma\_expansions.py} in the repository~\cite{Mukherjee_Fluid_limits_2026} checks all three symbolically.
\end{proof}

Thus $\sigma=0$ is a one-sided local minimum for \MRG{} within each of these two families.
For $S_n(\sigma)$, this endpoint is the Bomb graph.
For $S^K_n(\sigma)$, it is the Bomb graph with cliques added to both hub classes, which also has limiting ratio $2/3$.
A small positive satellite density improves the performance of \MRG{} in either family.
For example, $\rho_S(1/4)\approx0.6873$.
Numerically, the ratio for $S_n(\sigma)$ crosses $2/3$ near $\sigma=0.848$ and $\sigma=1.758$, and lies below it between these values.
For $S^K_n(\sigma)$, the crossings are near $\sigma=0.492$ and $\sigma=1.750$.
Both minima occur near $\sigma=1$.
A local search that starts at $\sigma=0$ and accepts only improvements therefore cannot leave that point by making small changes in $\sigma$.
For \RDO{}, every density in $0<\sigma<8/5$ gives a ratio below $2/3$.
This difference is consistent with the Double-Bomb having been found for \RDO{} first.

\paragraph{Numerical evidence for smaller constants:}
Local search over templates finds instances whose estimated ratios are about $0.6555$ for \MRG{} on general graphs (\cref{fig:candidates}(a)), $0.6603$ for \MRG{} on bipartite graphs (\cref{fig:candidates}(b)), and $0.6431$ for \RDO{} on bipartite graphs (\cref{fig:candidates}(c)).
The first adds a second layer of hubs to an asymmetric version of $S^K_n$, the second attaches pendant edges to the hubs of one side and to the antennas of the other side of a bomb graph, and the third adds two small gadgets to the Double-Bomb.
These are numerical estimates, and the specifications and convergence data are in the repository~\cite{Mukherjee_Fluid_limits_2026}.
Proving the corresponding bounds would require checking the count-chain and drift conditions, controlling the trajectory through any changes of phase, and justifying the terminal contribution.
We do not have closed forms for these trajectories.
Rigorous numerical bounds on the ODE solutions could be one way to complete the analysis, but we leave this as future work.

\paragraph{Open problems:}
\begin{enumerate}
\item For \MRG{} on general graphs, the gap is $[0.531,0.6569]$, and our numerical evidence suggests an upper bound near $0.6555$ (\cref{fig:candidates}(a)).
The lower bound is inherited from \RDO{}, whose approximation ratio is at most $F(w^*)=0.594531\ldots$ by \cref{thm:rdo-general}.
Thus an analysis that establishes only a guarantee for every fixed preference profile cannot prove a larger lower bound for \MRG{}.
Pushing the constant beyond $F(w^\ast)$ requires using the randomness of its preferences.
\item For \RDO{} on bipartite graphs, the gap is $[0.639,31/48]$, and our numerical search suggests an upper bound near $0.6431$ (\cref{fig:candidates}(c)).
Determining the exact ratio remains open.
\item It would be interesting to extend the toolkit to Ranking, FRanking and UUR~\cite{CCWZ18,DY26}, and to other vertex-iterative greedy problems.
In Ranking the preference order is the decision order, and in FRanking the decision order is adversarial.
In UUR the two orders are independent, but all vertices share the same preference order.
These dependencies are not captured by the counts used here, so the extensions would require additional state information and new arguments.
\end{enumerate}

\begin{acks}
This work is supported by the Center of Innovation for Sustainable Quantum AI (JST Grant Number JPMJPF2221, Japan).
The author thanks Vorapong Suppakitpaisarn for several helpful discussions, and Rex West and Yonghao Yue for several helpful comments that improved the presentation of the manuscript.
\end{acks}

\section*{AI Methodology Statement}
The author used Claude (Anthropic) and GPT (OpenAI) to assist with this work. The author formulated the research questions, directed the investigation, and selected the starting constructions and methodological references. In particular, the author supplied the Double-Bomb construction of Tang, Wu, and Zhang~\cite{TWZ23} and identified Mastin and Jaillet’s analysis of greedy matching on \(G(n,n,p)\)~\cite{MJ13} as a model for the application of Wormald’s differential equation method.

Under this direction, Claude assisted with developing and refining candidate constructions, derivations, and proof arguments, and wrote the code for template evaluation, searches, and simulations. Claude and GPT also assisted with the literature review, identifying potentially missing citations, checking mathematical arguments, and preparing preliminary manuscript drafts.

The author reviewed and revised the mathematical arguments, checked the computational results and references, and rewrote the manuscript for submission. The author takes full responsibility for the correctness, originality, and presentation of the work.
The code is available at \url{https://github.com/Potla1995/MRGUpperBd}~\cite{Mukherjee_Fluid_limits_2026}.

\bibliographystyle{alpha}
\bibliography{main_stoc27}

\end{document}